\documentclass[aps,pra,twocolumn,superscriptaddress,nofootinbib]{revtex4-2}
\usepackage{amssymb,graphicx,color}
\usepackage[intlimits]{amsmath}
\usepackage[english]{babel}
\usepackage[colorlinks]{hyperref}
\hypersetup{
colorlinks=blue,
linkcolor=blue,
filecolor=blue,
urlcolor=blue,
citecolor=blue
}
\pdfoutput=1
\usepackage[normalem]{ulem}
\usepackage{comment}
\usepackage{cprotect}
\usepackage{ragged2e}
\usepackage{enumitem}
\usepackage{amsmath}
\usepackage{mathtools}
\usepackage{amsmath}
\usepackage{amsfonts}
\usepackage{amssymb}
\usepackage{siunitx}
\usepackage{verbatim}
\usepackage{hyperref} % Required for inserting clickable links.
\usepackage{natbib} % Required for APA-style citations.
\usepackage{amsthm}
\usepackage{amsfonts}
\usepackage{braket}
\usepackage{float}
\usepackage{blkarray}
\usepackage{physics}
\usepackage{qcircuit}
\usepackage{multirow}
\usepackage{tikz}
\usetikzlibrary{calc}
\usetikzlibrary{matrix}
\usepackage{booktabs}
\usepackage{float}
\usepackage[ruled,vlined,linesnumbered]{algorithm2e}
\usepackage{algpseudocode}
\usepackage{tikz}
\usetikzlibrary{shapes.geometric, positioning}
\usepackage{pgfplots}
\usepgfplotslibrary{groupplots}
\pgfplotsset{compat=1.18}
\newcommand{\medium}{\fontsize{10pt}{12pt}\selectfont}
\begin{document}

\preprint{APS/123-QED}

\title{Genetic optimization of ansatz expressibility for
enhanced variational quantum algorithm
performance}% Force line breaks with \\

\author{Manish Mallapur}
\email{manish21@iiserb.ac.in}

\author{Ronit Raj}
\email{ronit21@iiserb.ac.in}

\author{Ankur Raina}
\email[Corresponding author: ]{ankur@iiserb.ac.in}

\affiliation{
    Department of Electrical Engineering and Computer Science, Indian Institute of Science Education and Research, Bhopal, 462066, Madhya Pradesh, India
}

% \author{Manish Mallapur}
%  \altaffiliation[Also at ]{Physics Department, XYZ University.}%Lines break automatically or can be forced with \\
%  \email{Second.Author@institution.edu}
% \affiliation{%
%  Authors' institution and/or address\\
%  This line break forced with \textbackslash\textbackslash
% }%

% \collaboration{MUSO Collaboration}%\noaffiliation

%  \homepage{http://www.Second.institution.edu/~Charlie.Author}
% \affiliation{
%  Second institution and/or address\\
%  This line break forced% with \\
% }%
% \affiliation{
%  Third institution, the second for Charlie Author
% \affiliation{%
%  Authors' institution and/or address\\
%  This line break forced with \textbackslash\textbackslash
% }%

% \collaboration{CLEO Collaboration}%\noaffiliation

\date{\today}% It is always \today, today,

\begin{abstract}
\textcolor{black}{Variational quantum algorithms have emerged as a leading paradigm that extracts practical computation from near-term intermediate-scale quantum devices, enabling advances in quantum chemistry simulations, combinatorial optimization, and quantum machine learning.
However, the performance of variational quantum algorithms is highly sensitive to the design of the ansatze. 
To be effective, ansatze must be expressive enough to capture target states but shallow enough to be trainable.
We propose a genetic algorithm-inspired framework for designing ansatze that achieve high expressibility while maintaining shallow depth and low parameter count. 
Our approach evolves ansatze through mutation and selection based on an expressibility metric. 
The circuit generated by our framework consistently demonstrates high expressibility at any target depth and performs comparably to traditional ansatz design approaches. 
This work presents a problem-agnostic, scalable solution for ansatz design, producing expressive, low-depth circuits that need to be designed only once and can serve a wide range of applications.}
% \begin{description}
% Secondary publications and information retrieval purposes.
% You may use the \texttt{description} environment to structure your abstract;
% use the optional argument of the \verb+\item+ command to give the category of each item. 
% \end{description}
\end{abstract}

\maketitle

%\tableofcontents

\section{Introduction}\label{sec1}

Variational quantum algorithms \cite{Cerezo2021} stand at the forefront of quantum computing's transition from theoretical promises to practical applications. 
In the noisy intermediate-scale quantum (NISQ) era \cite{preskill2018quantum,bharti2022noisy}, where quantum processors are characterized by their limited qubit counts and susceptibility to noise, variational quantum algorithms (VQAs) emerge as the leading algorithmic framework \cite{bergholm2018pennylane} for practical quantum computing applications. 
The primary advantage of VQAs is their ability to run shallow-depth circuits, which is crucial in NISQ devices, given their low coherence times and noise accumulation with an increase in circuit depth. 
The hybrid nature of this algorithm delegates the optimization process to classical computers while using the quantum computer for state preparations and state measurements. VQAs have demonstrated their practical utility across a range of applications that align with the capabilities of NISQ devices. 
 For example, the QAOA \cite{farhi2014quantum} has been employed to solve combinatorial optimization, including its similar variants like the Max-Cut problem \cite{wang2018quantum} and other constraint satisfaction problems \cite{lin2016performance}, error correction \cite{johnson2017qvector, bhattacharyya2023quantum}, and for solving linear systems of equations \cite{harrow2009quantum}. Additionally, Variational Quantum Eigen solver (VQE) \cite{peruzzo2014variational} is a prominent algorithm for finding ground state energies of molecular systems. 

However, the choice of ansatz determines the performance of VQAs \cite{Hubregtsen2021}. This impact arises from an ansatz's ability to explore the appropriate solution space, which determines how close the algorithm can approximate the true solution to the given problem. The performance of the ansatz depends on balancing the competing requirements. On the one hand, the expressibility of an ansatz is crucial for representing the complex and wide range of states required for accurate solutions. The trainability of the ansatz is also crucial for the algorithm's convergence. An ansatz with a huge number of quantum gates is supposedly highly expressible, but has poor trainability due to increased depth and parameter count. Several works have explored automated designs of ansatze to improve the performance of VQAs.  These quantum architecture search methods aim to optimize variational circuit design, often through evolutionary strategies like genetic algorithms and gate-level optimization \cite{williams1998automated,lamata2018quantum,huang2022robust}. Reinforcement learning based quantum architecture search explores these quantum architectures through policy-driven agents \cite{kuo2021quantum}. Predictor-based QAS \cite{zhang2021neural, he2023gnn} and SuperCircuit frameworks \cite{du2022quantum,wang2022quantumnas} have gained attention due to improved scalability factors, but they still pose a high classical computational cost.
The challenge is to find a balance between expressibility and trainability of the ansatz,  keeping the scalability in mind.

\begin{figure}[htbp]
\centering
\includegraphics[width=1\linewidth]{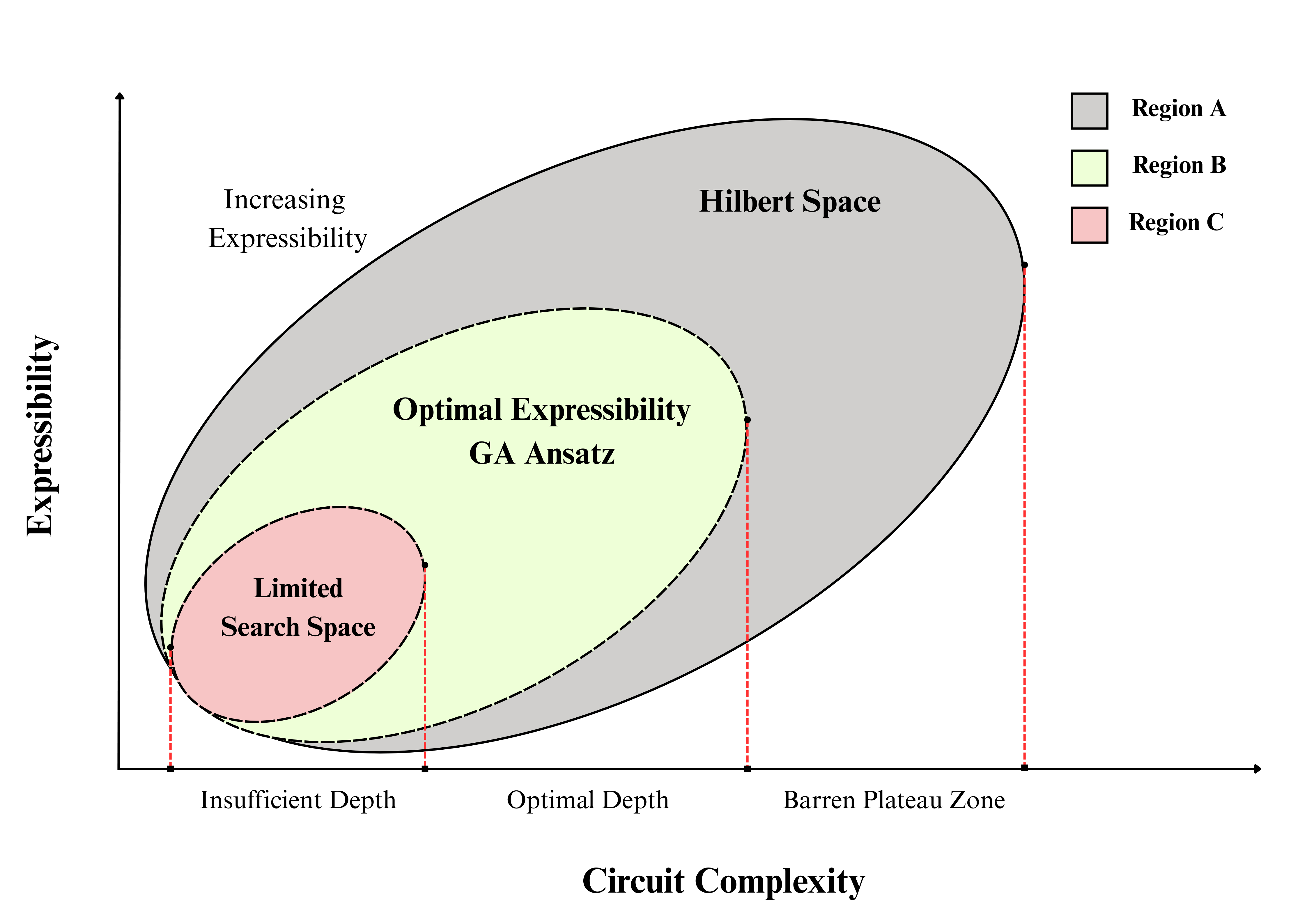}
\caption{\ There is a trade-off between circuit complexity and expressibility in variational quantum circuits. The diagram shows three important areas in the expressibility-complexity landscape. Region C (red) shows a small search space because the circuit depth is not deep enough, which makes it hard to get close to the target states. The green area in Region B shows the best expressibility zone that our genetic algorithm (GA)-generated ansatz can find. It does this by choosing the best gate configurations at the right depths to balance expressibility and trainability. At high levels of expressibility and complexity, Region A (gray) shows the entire Hilbert space that can be reached. Circuits suffer from the barren plateau phenomenon, which makes them hard to train. Our depth-aware GA framework moves around these areas in real time to create resource-efficient ansatze that are highly expressive and converge quickly across a wide range of problem sizes.}
\label{fig:sample}
\end{figure}

Our work introduces a genetic algorithm-based ansatz design framework \cite{Creevey2023,sunkel2023ga4qco} that addresses the limitations of current ansatz design schemes, which are often fixed and greedy in nature. Our method leverages the global search capabilities of evolutionary algorithms to explore the vast circuit design space. We guide the search towards circuits that maximize expressibility while operating on a fixed depth.
In other words, we use a classical algorithm as a pre-processing step before using variational principles on an NISQ computer.
We introduce depth-aware circuit optimization strategies where, at each depth, the framework identifies optimal configurations that balance gate efficiencies and expressibility. 
This systematic approach provides us a practical way to navigate the expressibility-trainability \cite{Holmes_2022} tradeoff, which is inherent in variational quantum algorithms. To validate the effectiveness of our framework, we apply the resulting ansatze to one of the most prominent VQA applications: finding the ground state energy of complex quantum systems using VQE. 
We benchmark their performance on a range of molecular Hamiltonians (\ref{subsec:mol}) as well as spin Hamiltonians (\ref{subsec:tfim}), demonstrating practical utility for both quantum chemistry and condensed matter problems. Through this work, we demonstrate the potential of genetic algorithms as a ansatz design tool capable of generating resource-efficient, highly expressible circuits across various problems, sizes and depths.

This paper is organized as follows. In Section \ref{sec2}, we briefly review the fundamentals of the VQE, including the critical tradeoff between ansatz expressibility and trainability. Section \ref{subsec: related-work} talks about the related works and established ansatz designs like UCCSD and ADAPT-VQE, which also serve as benchmarks. We then provide an overview of genetic algorithms in section \ref{sec5 : A} and the evolutionary optimization techniques that form the basis of our method.  Our novel genetic algorithm-based framework for designing highly expressible quantum circuits is detailed in Section \ref{sec5 : B}. \textcolor{black}{In Section \ref{sec6}, we present a comprehensive analysis of our framework, where we evaluate the impact of  various parameters, benchmarking the generated ansatze on various Hamiltonians.} Finally, we discuss the implications of our findings in Section \ref{sec7} and present our conclusion in Section \ref{sec8}.

\vspace{-0.8cm}
\section{Variational Quantum Eigensolver (VQE)}\label{sec2}

The variational principle in quantum mechanics states that for any normalized trial wavefunction, the expectation value of the Hamiltonian operator will always give a value greater than the true ground state energy of the system. This principle has become the basis for many variational approaches, serving as a powerful theoretical framework for approximating quantum systems. Variational quantum algorithms tackle optimization problems by encoding the solution into a cost function $C(\theta)$. A parameterized quantum circuit (ansatz $U(\theta)$) is then proposed and a classical optimizer updates $\theta$ to minimize the cost function, where the minima corresponds to the problem solution. This ansatz is trained iteratively in a hybrid quantum-classical manner to solve the optimization task.

The cost function defines the solution landscape over which the hybrid loop iterates, directly influencing algorithm performance. It must be faithful (minima correspond to problem solutions), feasible to compute on quantum hardware, not classically trivial, operationally meaningful (lower cost means better solution), and smooth enough for optimization. 

While the cost function provides direction for optimization, the variational ansatz defines the landscape to be explored. The ansatz acts as a template that dictates expressibility and resource efficiency. Expressibility is the ansatz's ability to span arbitrary space with respect to target states, directly impacting its ability to reach the desired solution. The ansatz structure fundamentally dictates both expressibility (how well it represents states) and trainability (how easily parameters can be optimized). However, practical issues like barren plateaus and local minima can hinder performance despite high expressibility \cite{du2022quantum}. Higher expressibility allows better representation of quantum states, which is necessary for accurate solutions, but trainability determines how easily circuit parameters can be optimized for convergence. As shown in \cite{Holmes_2022}, ansatze with many quantum gates become highly expressive, but noise accumulation often results in poor trainability. \textcolor{black}{The goal is to produce sufficiently expressible ansatz that are resource efficient (i.e., have shallow depths) while keeping the circuits trainable.}

\subsection{System Hamiltonians of interest}
To evaluate our ansatz circuits, we tested them on several molecular systems and compared results with state-of-the-art designs (UCCSD and ADAPT-VQE). The classical optimization was performed using Adam optimizers. We also tested on a non-molecular quantum system to check generalization.
\vspace{-0.5cm}
\subsubsection{Molecular Hamiltonians}

We used H$_2$, LiH, BeH$_2$, and H$_2$O. Each molecular system is described by an electronic Hamiltonian in second quantization \cite{helgaker2000molecular}:

\begin{align}
\hat{H} = \sum_{pq} h_{pq} \, a_p^\dagger a_q + \frac{1}{2} \sum_{pqrs} h_{pqrs} \, a_p^\dagger a_q^\dagger a_r a_s,
\end{align}

Where $a_p^\dagger$ and $a_p$ are the fermionic creation and annihilation operators that add or remove electrons in specific spin-orbitals, $h_{pq}$ are one-electron integrals (kinetic energy and nuclear attraction), and $h_{pqrs}$ are two-electron integrals (electron-electron repulsion).

These integrals are computed from molecular orbital data (molecular positions, bond lengths, etc.) using the STO-3G minimal basis set \cite{hehre1972sto3g}. Since quantum computers work with qubits rather than fermions, we map this fermionic Hamiltonian to a qubit Hamiltonian using the Jordan-Wigner transformation \cite{jordan1928pauli, seeley2012bravyi}.

Our benchmark molecules \cite{peruzzo2014variational} are: H$_2$ (simple two-electron system, four qubits), LiH (4 electrons, 12 qubits), BeH$_2$ (linear molecule, 6 electrons, 14 qubits), and H$_2$O (bent molecule, 10 electrons, 14 qubits).

\subsubsection{Non-Molecular Hamiltonian: Transverse field Ising model}

We also tested our approach on the transverse field Ising model (TFIM), a one-dimensional chain of spin-$\frac{1}{2}$ particles with nearest-neighbor interactions and uniform transverse magnetic field. The TFIM Hamiltonian \cite{sachdev2011quantum, dutta2015quantum} is:
\begin{align}
\hat{H}_{\text{TFIM}} = -J \sum_{\langle i, j \rangle} Z_i Z_j - h \sum_{i} X_i,
\end{align}

where $Z_i$ and $X_i$ are Pauli operators acting on qubit $i$, and $\langle i,j \rangle$ refers to nearest-neighbor interactions. The $Z_i Z_j$ term represents spin-spin interactions between neighbouring qubits, and the $X_i$ term represents an external transverse magnetic field. \textcolor{black}{We use $J = 1$ and $h = 1$ in our experiments.}
\vspace{-0.5cm}
\subsection{Expressibility}

 Expressibility characterizes a PQC's ability to generate diverse quantum states across the Hilbert space. Several approaches to defining and measuring expressibility have been proposed in the literature \cite{sim2019expressibility,nakaji2021expressibility,schuld2021effect,holmes2022connecting}. In this work, we adopt the fidelity-based method introduced by \cite{sim2019expressibility}, as it is an operationally meaningful and practically estimable measure of how well a circuit can explore the Hilbert space. A circuit is said to be highly expressible if it generates states that can approximate the uniform Haar distribution. 

In practice, expressibility is estimated by sampling pairs of random parameters $\theta$ and $\phi$, drawn uniformly from $[0,2\pi]$, preparing the corresponding states $\psi(\theta)$ and $\psi(\phi)$, and computing their fidelity:
\begin{equation}
    F = |\langle \psi(\theta) | \psi(\phi) \rangle|^2.
\end{equation}

Repeating this procedure many times yields an empirical distribution $\hat{P}_{\text{PQC}}(F)$ of fidelities. This is compared against the known fidelity distribution of Haar-random states,
\begin{equation}
    P_\text{Haar}(F) = (N-1)(1-F)^{N-2},
\end{equation}

\noindent where $N = 2^n$ is the Hilbert space dimension. 
Sim \emph{et al.} originally proposed using the Kullback–Leibler (KL) divergence between $\hat{P}_{\text{PQC}}(F)$ and $P_\text{Haar}$ distributions to quantify expressibility.

In our work, we use the Jensen-Shannon divergence, a symmetric and bounded alternative to the KL divergence.

        \begin{equation}
        E = D_{\text{JSD}}(\hat{P}_{\text{PQC}}(F) \parallel P_{\text{Haar}}(F)). \label{eq:exp}
       \end{equation}

A lower divergence value indicates higher expressibility, reflecting a closer approximation to the Haar distribution.

\vspace{-0.5cm}
\subsection{Related work} \label{subsec: related-work}

% To benchmark the performance of our ansatz, we compare it against two widely-used VQE ansatz:

Contemporary VQE ansatz design spans a broad spectrum, including hardware-efficient circuits tailored to device constraints \cite{Kandala2017}, problem-specific constructions such as the Hamiltonian variational ansatz \cite{Wecker2015}, and chemically motivated approaches that encode the molecular electronic Hamiltonian, notably Unitary Coupled Cluster with Singles and Doubles (UCCSD) \cite{https://doi.org/10.1002/qua.21198} and adaptive schemes like ADAPT-VQE \cite{Grimsley2019}.

Unitary Coupled Cluster with Singles and Doubles (UCCSD) and adaptive schemes like ADAPT-VQE have undergone numerous variations \cite{Ollitrault2024, Mizukami2020, Lee2019, Evangelista2022, Tang2019, Shkolnikov2023, Yordanov2021, Wang2025, Smith2024}. We focus on the original UCCSD and ADAPT-VQE since we worked mainly on molecular Hamiltonians to compare them against established molecular ansatz designs.

\noindent
\begin{enumerate}
\item UCCSD: Unitary Coupled Cluster with Singles and Doubles is a chemically inspired ansatz that adapts the accurate coupled cluster method from quantum chemistry into a unitary form suitable for quantum computation. The method constructs the ansatz by applying a unitary transformation to a reference Hartree–Fock state:

\begin{align}
\ket{\Psi(\boldsymbol{\theta})} = e^{T - T^\dagger}\ket{\Phi_0},
\end{align}

In the following, let \(\ket{\Phi_0}\) denote the reference Hartree–Fock state; \(\ket{\Psi(\boldsymbol{\theta})}\) and \(\ket{\psi(\boldsymbol{\theta})}\) the parameterized ansatz wavefunctions; \(\boldsymbol{\theta}\) the vector of variational parameters; \(n\) the number of spin–orbitals; \(a_p^\dagger\) and \(a_q\) the fermionic creation and annihilation operators; and \(\hat{H}\) the molecular Hamiltonian. For UCCSD, the excitation operator \(T\) is decomposed as \(T=T_1+T_2\)
 which includes single excitation operators:
\begin{align}
T_1 = \sum_{i,a} \theta_i^a\, a_a^\dagger a_i,
\end{align}
and double excitation operators:
\begin{align}
T_2 = \sum_{i<j,\,a<b} \theta_{ij}^{ab}\, a_a^\dagger a_b^\dagger a_j a_i.
\end{align}
with \(\theta_i^a\) and \(\theta_{ij}^{ab}\) the amplitude parameters for single and double excitations from occupied orbitals \(i,j\) to virtual orbitals \(a,b\), respectively.
Since direct implementation of the unitary exponential is challenging, it is approximated using a first-order Trotter–Suzuki decomposition \cite{doi:10.1126/science.273.5278.1073} and mapped to qubits via fermion-to-qubit encodings. At the same time, UCCSD provides accurate results for small molecules, its \(\mathcal{O}(n^3)\) scaling in parameterized gates with the number of spin–orbitals \(n\) limits scalability to larger systems.

\medskip

\noindent

\item ADAPT-VQE: Adaptive Derivative-Assembled Pseudo-Trotter Ansatz Variational Quantum Eigensolver is an iterative ansatz construction method that builds circuits one gate at a time based on energy gradients. 
The parameterized state is 

\begin{align}
\ket{\psi(\boldsymbol{\theta})} = U_N(\theta_N)U_{N-1}(\theta_{N-1})\cdots \ket{\Phi_0},
\end{align}
where each unitary 

\begin{align}
U_k(\theta_k) = e^{-i\theta_k \hat{A}_k},
\end{align}
uses operators from a fermionic excitation pool. The algorithm iteratively computes gradients

\begin{equation}
    g_j = \left| \bra{\psi(\boldsymbol{\theta})} [\hat{H},\hat{A}_j] \ket{\psi(\boldsymbol{\theta})} \right|, 
\end{equation}

and selects the operator with maximum \(g_j\), appends its unitary to the ansatz, and optimizes all parameters until convergence. Although ADAPT-VQE can produce shallower circuits than fixed ansatze, computing gradients for all operators at each iteration incurs significant classical overhead.
\end{enumerate}

\vspace{-0.5cm}
\section{The ansatz design problem}\label{sec5}
% \textcolor{red}{Designing the right ansatz is a very important task in the usage of variational algorithms.
% One is required to balance expressibility, trainability, circuit depth and circuit width to name a few while optimizing a cost function. 
% We highlight below our genetic algorithm-driven pre-processing unit that generates a highly expressible ansatz for circuits of varying complexity.} 

The performance of VQAs is critically dependent on the design of the parameterized quantum circuit, or ansatz. On one hand, it requires sufficient expressibility to explore the relevant regions of the Hilbert space and accurately represent the target quantum state. On the other hand, increasing circuit complexity to boost expressibility often introduces significant challenges for Noisy Intermediate-Scale Quantum (NISQ) hardware, including increased susceptibility to noise and the emergence of optimization-hindering phenomena like barren plateaus. \textcolor{black}{The task is therefore to find resource-efficient circuits that do not sacrifice performance.} Given the vast, high-dimensional, and non-convex nature of the search space for circuit architectures, heuristic methods offer a promising path forward. In this section, we introduce an approach based on a genetic algorithm to systematically search for and evolve ansätze that balance these competing requirements.
\vspace{-0.5cm}
\subsection{Genetic Algorithm} \label{sec5 : A}

\begin{figure*}[htbp]
\centering
\includegraphics[width=0.75\linewidth]{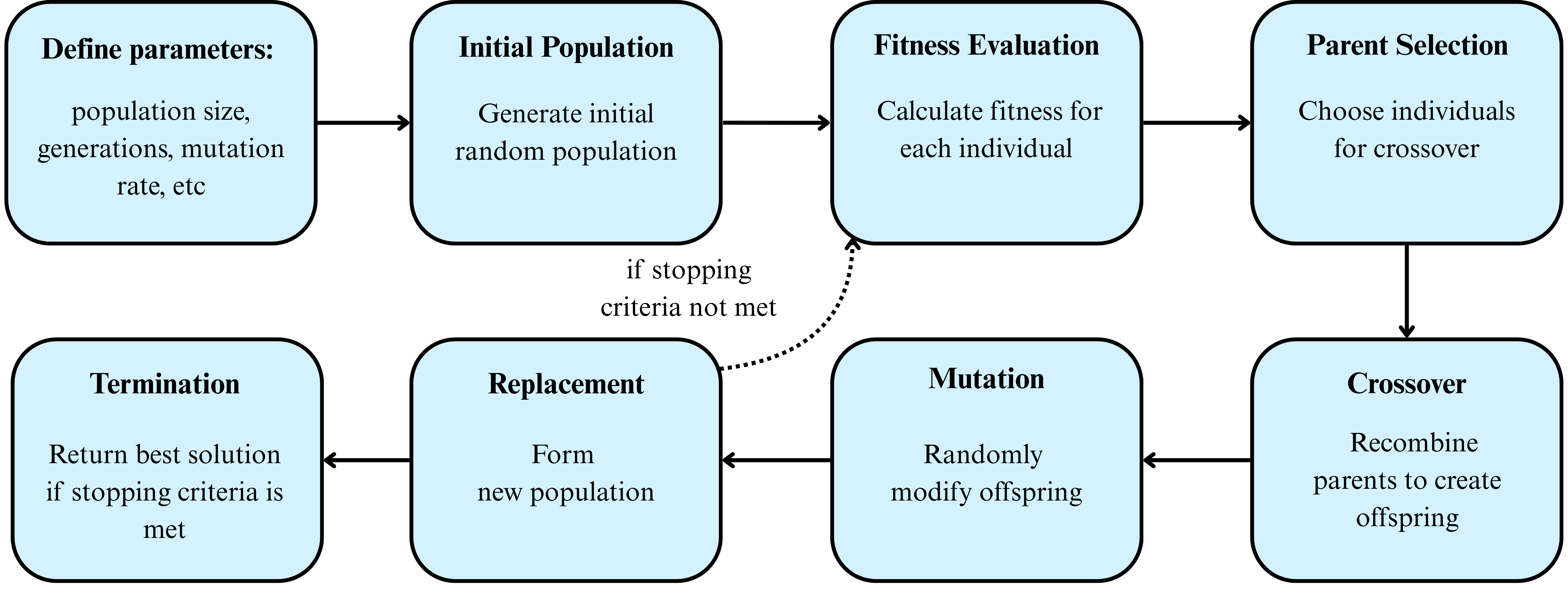}
\caption{Overview of the general Genetic Algorithm: parameter initialization, generation of the initial population, iterative fitness evaluation, selection of parents, crossover to produce offspring, mutation for genetic diversity, population replacement, and termination upon meeting the stopping criteria.}
\label{fig:gageneral}
\end{figure*}

Genetic algorithms \cite{Goldberg1988, Mirjalili2019} are a heuristic evolutionary optimization algorithm inspired by natural selection, where in a population, only those individuals with the most suitable genetic traits are able to propagate and pass their characteristics to subsequent generations. This translates into a powerful computational framework that has proven remarkably effective in tackling complex optimization landscapes. Genetic algorithms \cite{zhang2023evolutionary} are found to work well in many cases where traditional optimization techniques struggle, such as problems with non-convex search spaces, scenarios requiring non-gradient optimization, or when working with high-dimensional data where classical methods often get trapped in local optima. For our specific case, the genetic algorithm is able to optimize the ansatz for expressibility given the constraints like the number of qubits and maximum circuit depth, essentially searching through the vast space of possible quantum circuit configurations to find those that best balance performance and resource requirements. Genetic algorithms emulate the process of evolution through biologically inspired operations like selection (where fitter individuals have higher chances of reproduction), crossover (where genetic material is exchanged between parent solutions to create offspring), and mutation (where small random changes are introduced to maintain diversity and prevent premature convergence) to produce high-quality solutions to the optimization problem. The iterative nature of this process allows the algorithm to gradually improve the population quality over successive generations, with each cycle potentially discovering better circuit designs that might be difficult to find through traditional optimization approaches. The general loop of a genetic optimization algorithm is as shown in Figure \ref{fig:gageneral}, which illustrates how these evolutionary principles work together to guide the search toward increasingly effective solutions.

\vspace{-0.5cm}

\subsection{Our ansatz design algorithm} \label{sec5 : B}

In this work, we develop a framework inspired by genetic algorithms to identify highly expressible circuits (ansatze) with limited circuit depth. This evolutionary approach is well-suited to this task, as we have to search through a non-convex and high-dimensional landscape of quantum circuit designs. Our approach encodes quantum circuits as individuals in the population, where each circuit's expressibility is treated as its fitness score. Through the iterative process of selection, crossover, and mutation, the population evolves toward circuits with higher expressibility.

% Outer list: A), B), C)...
\setlist[enumerate,1]{label=\Alph*)}

% Inner list: a), b), c)...
\setlist[enumerate,2]{label=\alph*)}

\begin{enumerate}
    \item Representation: Each individual in the population of the genetic algorithm (GA) is a unique quantum circuit ansatz. 
    The individual is represented as a sequence of quantum gates arranged layer-by-layer, along with a set of corresponding parameters for the parameterized gates (RX, RY, and RZ). We treat the ordered set of all the gates in the circuit as the genome of the individual, with each individual gate, along with its parameters, functioning as a gene. This genome-based encoding facilitates genetic operations such as crossover and mutation in the evolutionary process.

    \item Initialization: The population is initialized with random quantum circuits up to a specified depth. Each layer is built randomly, choosing gates from the predefined gate sets (gates = [RX, RY, RZ, H, I, and CNOT]). For gates with parameters, the parameters are uniformly sampled randomly from $[0,2\pi]$

    \item Fitness Evaluation: The fitness of an individual is defined by the expressibility of corresponding ansatze. Expressibility is measured by calculating the Jensen-Shannon divergence (JSD) between the distributions of the fidelities [calculated from random sampling of the circuit] and the Haar random distribution of the Hilbert space. A lower JSD value suggests higher expressibility, making the fitness inversely proportional to JSD.

    \item Selection, Crossover, and Mutation: 
    \begin{enumerate}
        \item Selection: Individuals are ranked on the basis of their expressibility from most expressible (lowest JSD) to least expressible (highest JSD) and choose the top `$k$' individuals as parents for the next generations.
        \item Crossover:  In generating new offspring, the circuits are split between layers and recombined. This introduces the structural variations in the offspring compared to their parents.

        \item Mutation: A random subset of gates is chosen to be mutated by replacing a gate with a random gate or changing the order of gates.
        
    \end{enumerate}

After several generations, the genetic algorithm framework converges towards ansatze with maximized expressibility for a given depth and qubit count.

\end{enumerate}

\begin{figure*}[htbp]
\centering
\includegraphics[width=0.75\linewidth]{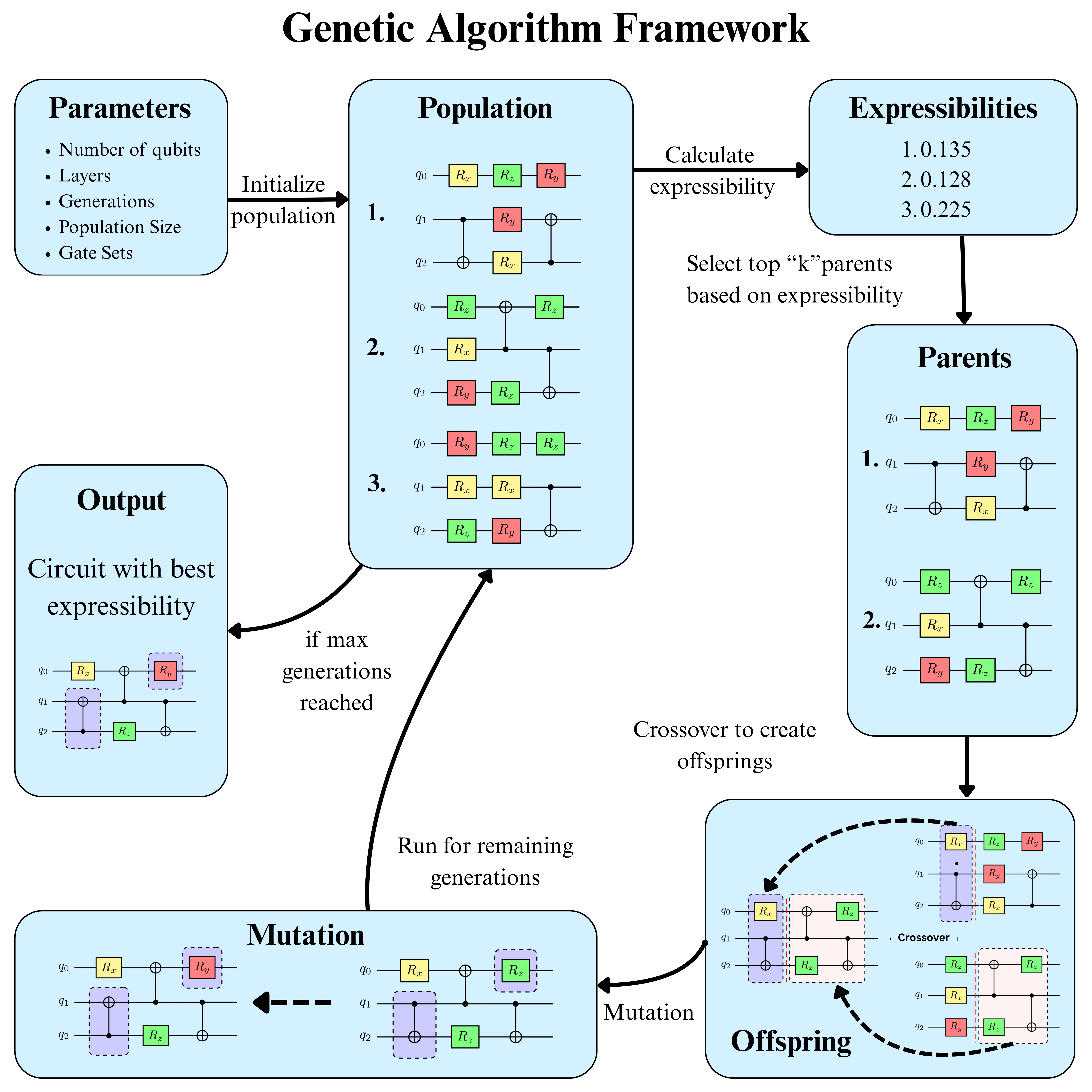}
\caption{Genetic algorithm framework for ansatz optimization. A population of random quantum circuits is initialized and evaluated using expressibility as the fitness metric. Top-performing circuits are selected as parents for crossover, generating offspring by recombining gate sequences. 
Offspring undergo mutation to maintain population diversity. The process iterates over generations, yielding an ansatz with maximal expressibility under depth and qubit constraints.}
\label{fig:GA-framework}
\end{figure*}

The following section is a step-by-step implementation of our genetic algorithm to optimize for the expressibility of the ansatz for given constraints like number of qubits, circuit depth, or number of parameters, as shown in Fig. \ref{fig:GA-framework}. 

\begin{enumerate}
    \item Constraints for optimization:
    \begin{enumerate}
        \item Number of qubits 
        \item Number of layers 
        \item Population size 
        \item Number of generations 
        \item Mutation rate 
        \item Number of samples for expressibility calculation 
        \item Available quantum gates: {RX, RY, RZ, H, CNOT}
    \end{enumerate}

    \item Initialize the population: Generate population size quantum circuits, each with a number of layers. Gates are randomly chosen from the available set. For parameterized gates (RX, RY, RZ), assign random parameters in the range $[0, 2\pi]$.

    \item Evaluate expressibility for each circuit:
    \begin{enumerate}
        \item Sample two random parameter sets and compute the corresponding quantum states.
        \item Compute fidelity $F = |\langle \psi(\theta) | \psi(\phi) \rangle|^2$.
        \item Construct a histogram of fidelities and compare it to the Haar-random distribution using JSD divergence as described in Eq. \ref{eq:exp} :
        \[
        E = D_{\text{JSD}}(P_{\text{circuit}}(F) \parallel P_{\text{Haar}}(F)).
        \]
        \item A lower expressibility score ($E$) indicates a better circuit.
    \end{enumerate}

    \item Select parents for the next generation: Rank circuits based on their expressibility scores and select the top $k$ (e.g., 5) circuits.

    \item Generate offspring via crossover and mutation:
    \begin{enumerate}
        \item Crossover: Select two parent circuits, and their genes are crossed to produce the offspring. This process can be done in two ways:
        \begin{enumerate}
            \item Single Point Crossover: We choose a random crossover point (layer index), and combine parts from both parents to form a child circuit.
            \item  \texttt{N} point crossover: Here, instead of choosing a single point to split the two parents, \texttt{N} random points are chosen on both parents, thus dividing them into \texttt{N+1} parts that are then alternatively joined to give the offspring.

            \begin{figure*}[htbp!]
            
            \centering
            \includegraphics[width=0.95\linewidth]{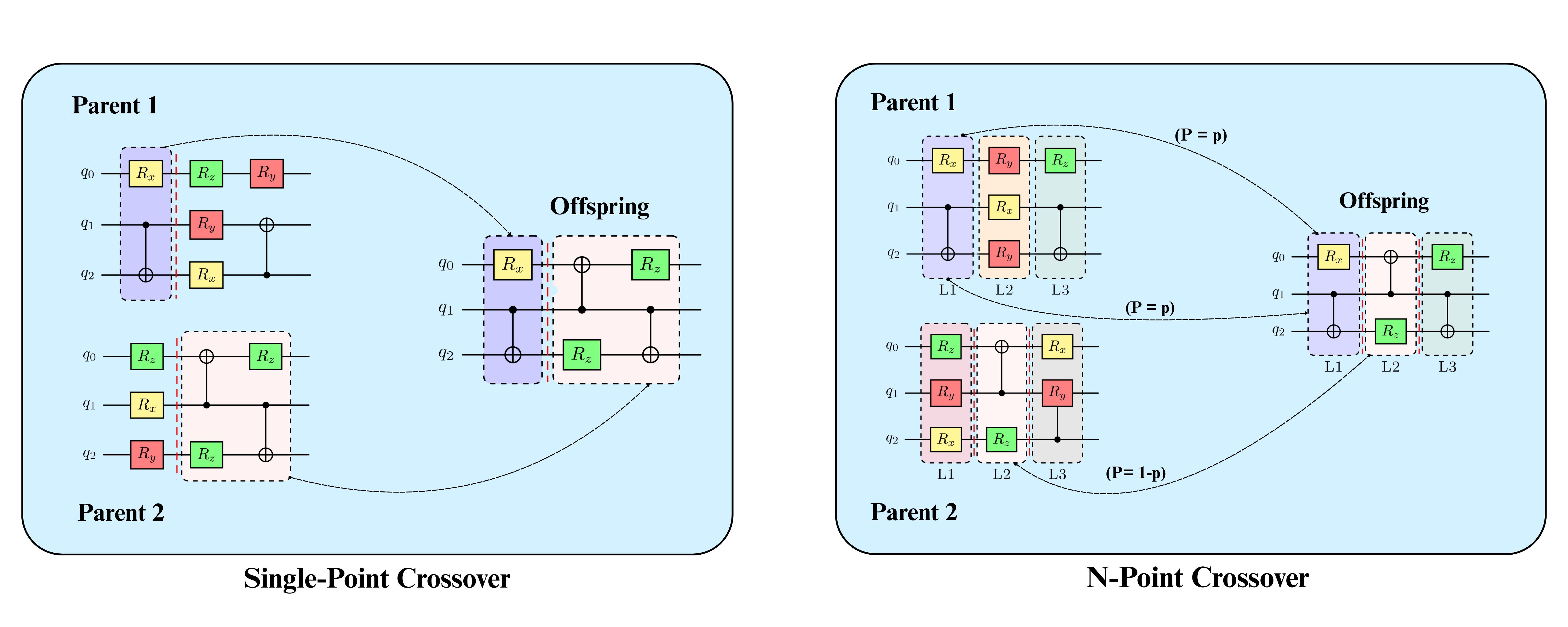}
            \caption{Single-point crossover. A single point splits the parent circuits, and the tail segments from both parents exchange to generate offspring that inherit partial gate sequences from both parents.
            $N$-point crossover. Multiple crossover points are selected across parent circuits, and segments are probabilistically swapped to create offspring that are diverse and preserve structural traits from both parents.}
            \label{fig:sample}
            \end{figure*}

        \end{enumerate}
        \item Mutation: With probability mutation rate, randomly modify a gate in the child circuit. These mutations are of two types:
        \begin{enumerate}
            \item Single qubit gate: If a single qubit gate is selected to be mutated, then it is replaced by another single qubit gate from the gate set.
            \item Two qubit gate (CNOT gate): If a CNOT gate is chosen to be mutated, then the control and target of the CNOT gate are swapped.
        \end{enumerate}
    \end{enumerate}

    \item Update population and iterate: Replace the old population with the new generation and repeat the process for the number of generations iterations.

\end{enumerate}

\begin{algorithm}[H]
\DontPrintSemicolon
\caption{The Ansatz design algorithm}\label{alg:main}
\KwIn{$n$ (qubits), $L$ (circuit depth), $P$ (population size), $S$ (samples), $G$ (generations), $k$ (parent count), $m$ (mutation prob.)}
\KwOut{Ansatz with optimal expressibility}
\textbf{Step 1: Initialize Population} \\
Create a population of $P$ random circuits with the given input constraints and compute their expressibility. (Alg.~\ref{alg:init})\;

\For{$g\leftarrow1$ \KwTo $G$}{
  \textbf{Step 2: Selection and Crossover} \\
  The best parents are selected on the basis of their expressibility. The parents are crossed and mutated to generate the offspring population using the current population. (Alg.~\ref{alg:cross})\;
  
  \textbf{Step 3: Population Update} \\
  Replace the current population with the offspring and track the best circuit. (Alg.~\ref{alg:update})\;
}
\Return{\textnormal{Best ansatz found over $G$ generations}}
\end{algorithm}

\subsection{Generalizability of the GA‐derived ansatz}
\label{sec:generalizability}

Instead of designing a new ansatz for each individual problem, we aim to build one general ansatz. We can use this circuit for different Hamiltonians, reducing the overhead of running the genetic algorithm every time we look at a new system. Once the circuit is designed, it works as an ansatz for different target Hamiltonians with some parameter optimization.

It lowers the classical overhead because it avoids running the genetic algorithm repeatedly, making the process simpler. It allows for faster experimentation as we can reuse the ansatz for different Hamiltonians.

Section~\ref{subsec:tfim} and~\ref{subsec:mol} show that the ansatz, which was not designed with a particular target Hamiltonian in mind, works well across different molecular and spin Hamiltonians.

\vspace{-0.5cm}
\section{Results}\label{sec6}

In this section, we provide a comprehensive analysis of the genetic algorithm (GA) framework developed to optimize ansatz expressibility. We systematically investigate the impact of various parameters, including the number of generations, circuit depth, gate set composition, and gate count, on the performance of the generated ansatze. Furthermore, we evaluate the real-world applicability of these GA-designed ansatze by benchmarking their performance on the variational quantum eigensolver (VQE) for both spin and molecular Hamiltonians. Our findings highlight key insights into designing efficient and effective ansatz for near-term quantum applications.
\vspace{-0.5cm}
\subsection{Genetic Algorithm (GA) framework}
To characterize the performance of our GA framework, we systematically varied its parameters. We start by analyzing the convergence of expressibility across generations to determine the optimal stopping point of the algorithm (\ref{subsubsec:expressibility convergence}). We then proceed to explore the relationship between circuit depth and expressibility, and identify the point of saturation where deeper circuits yield diminishing returns (\ref{subsubsec:expressibility saturation with circuit depth}). A central part of our analysis also investigates how different gate sets can impact expressibility (\ref{subsubsec:gate set}), resource requirement in terms of gate count (\ref{subsubsec:resource requirement}), and classical runtime (\ref{subsubsec:classical runtime}) of the framework to understand the tradeoffs involved in the ansatz design process.

\textcolor{black}{Before presenting the results, we first define the gate sets used throughout this work. The performance of the genetic algorithm and the resulting ansatz depends strongly on the choice of gates and their connectivity constraints. All subsequent analyses assume the gate set definitions summarized in Table \ref{table:gate_set_revised}.}

\renewcommand{\arraystretch}{1.5} % Controls row height
\begin{table}[h!]
\color{black}
\centering
\caption{
\textcolor{black}{Description of Gate sets A--I. Each gate set specifies the allowed single- and two-qubit operations, including nearest-neighbour CNOT connectivity and the presence of an initial Hadamard layer. An asterisk ($^*$) indicates that the CNOT gate is restricted to nearest-neighbor qubits. Gate sets marked with a double dagger ($^\ddagger$) include an initial layer of Hadamard (H) gates.}
}
\label{table:gate_set_revised}
\begin{tabular}{|c|l|}
\hline
\textbf{Set} & \textbf{Gates} \\
\hline
A & $\{ \mathrm{RX}, \mathrm{RY}, \mathrm{RZ}, \mathrm{H}, \mathrm{CNOT}^* \}$ \\
\hline
B & $\{ \mathrm{RX}, \mathrm{RY}, \mathrm{H}, \mathrm{CNOT}^*, \mathrm{I} \}$ \\
\hline
C & $\{ \mathrm{RX}, \mathrm{RY}, \mathrm{RZ}, \mathrm{H}, \mathrm{CNOT}, \mathrm{I} \}$ \\
\hline
D & $\{ \mathrm{RY}, \mathrm{RZ}, \mathrm{H}, \mathrm{CNOT}, \mathrm{I} \}$ \\
\hline
E & $\{ \mathrm{RY}, \mathrm{RZ}, \mathrm{CNOT}, \mathrm{I} \}$ \\
\hline
F$^\ddagger$ & $\{ \mathrm{RX}, \mathrm{RY}, \mathrm{RZ}, \mathrm{CNOT}, \mathrm{I} \}$ \\
\hline
G & $\{ \mathrm{RX}, \mathrm{RY}, \mathrm{RZ}, \mathrm{H}, \mathrm{CNOT}, \mathrm{I} \}$ \\
\hline
H$^\ddagger$ & $\{ \mathrm{RX}, \mathrm{RY}, \mathrm{RZ}, \mathrm{H}, \mathrm{CNOT}^* \}$ \\
\hline
I & $\{ \mathrm{RX}, \mathrm{RY}, \mathrm{RZ}, \mathrm{CNOT}, \mathrm{I} \}$ \\
\hline
\end{tabular}
\end{table}

\subsubsection{Number of generations}\label{subsubsec:expressibility convergence}
We began our analysis by determining the convergence behavior of the GA. As Fig.~\ref{fig:exp_vs_gen} illustrates, we observed the best expressibility scores for each gate set over ten generations. The plot (Fig.~\ref{fig:exp_vs_gen}) clearly shows that the GA framework finds more expressive ansatze over generations. For nearly all gate sets, the improvement saturates and plateaus around the tenth generation. This indicated that 10 generations provide near optimal expressibility without unnecessary computational overhead.

\begin{figure}
\centering
\color{black}
\begin{tikzpicture}
\begin{axis}[
    xlabel={Generations},
    ylabel={Expressibility},
    grid=major,
    legend style={at={(0.784,0.975)}, anchor=north, legend columns=3, font=\small},
      width=9.5cm,
      height=6cm,
    mark options={scale=0.8},
    ymin=0.04, ymax=0.10,
    ytick={0.04, 0.06, 0.08, 0.10},
    yticklabels={0.04, 0.06, 0.08, 0.10},
    % ----------------------------
    xtick={1,2,3,4,5,6,7,8,9,10}
]

% Plot data for gate sets A to I
\addplot+ [smooth, mark=*] coordinates {(1,0.06112) (2,0.06331) (3,0.05377) (4,0.05090) (5,0.04154) (6,0.05470) (7,0.04736) (8,0.04860) (9,0.04442) (10,0.05034)};
\addplot+ [smooth, mark=triangle*] coordinates {(1,0.06118) (2,0.05337) (3,0.05324) (4,0.05199) (5,0.04559) (6,0.04183) (7,0.05347) (8,0.04981) (9,0.05150) (10,0.05328)};
\addplot+ [smooth, mark=square*] coordinates {(1,0.07548) (2,0.05913) (3,0.06035) (4,0.06521) (5,0.05363) (6,0.05965) (7,0.05836) (8,0.04994) (9,0.05011) (10,0.05422)};
\addplot+ [smooth, mark=diamond*] coordinates {(1,0.09425) (2,0.08083) (3,0.07247) (4,0.07092) (5,0.06499) (6,0.06530) (7,0.06101) (8,0.06180) (9,0.05639) (10,0.05664)};
\addplot+ [smooth, mark=star] coordinates {(1,0.08160) (2,0.05947) (3,0.05829) (4,0.05184) (5,0.05538) (6,0.05586) (7,0.04884) (8,0.04588) (9,0.05439) (10,0.05121)};
\addplot+ [smooth, mark=triangle] coordinates {(1,0.06236) (2,0.06650) (3,0.06003) (4,0.05664) (5,0.05068) (6,0.05201) (7,0.05797) (8,0.05289) (9,0.05128) (10,0.05350)};
\addplot+ [smooth, mark=diamond] coordinates {(1,0.07434) (2,0.07202) (3,0.05898) (4,0.06058) (5,0.05535) (6,0.05498) (7,0.04718) (8,0.05811) (9,0.05523) (10,0.05781)};
\addplot+ [smooth, mark=otimes] coordinates {(1,0.05896) (2,0.06077) (3,0.05269) (4,0.05551) (5,0.04756) (6,0.04716) (7,0.04325) (8,0.04914) (9,0.05074) (10,0.04892)};
\addplot+ [smooth, mark=oplus] coordinates {(1, 0.0614) (2, 0.0492) (3, 0.0430) (4, 0.0503) (5, 0.0464)
(6, 0.0421) (7, 0.0476) (8, 0.0453) (9, 0.0510) (10, 0.0457)
};

\legend{A, B, C, D, E, F, G, H, I}

\end{axis}
\end{tikzpicture}
\caption{The plot shows the expressibility metric (y-axis) as a function of the GA framework generations (x-axis) for all gate sets (A-I) on a four-qubit system. A lower metric value indicates better expressibility. All curves show improvement in expressibility through generations, which consistently saturates by generation 10. We additionally verify this by extending the GA framework to 20 generations (see Figure \ref{fig:20-gen} in the appendix), which shows that exprressibility gains are minimal, confirming 10 generations as sufficient optimization duration. }
\label{fig:exp_vs_gen}
\end{figure}

\subsubsection{Circuit depth} \label{subsubsec:expressibility saturation with circuit depth}
After establishing the necessary generation convergence and we move to investigate the relationship between expressibility and circuit depth. Fig.~\ref{fig:exp_saturation} plots the expressibility for gate set A across systems of 4, 6, 10, 14 qubits. A consistent trend is observed across all systems: expressibility significantly improves as circuit depth increases, but only up to a saturation point. The saturation indicates that adding more layers yields diminishing results beyond a certain depth. \textcolor{black}{Our results suggest that this optimal depth scales linearly with the number of qubits, where optimality is defined empirically in terms of expressibility saturation, $ optimal \ depth\approx 2n$.} This finding is significant as it demonstrates that high expressibility can be achieved without requiring deeper circuits.

\begin{figure}
\color{black}
\centering
\vspace{2mm}
\begin{tikzpicture}
\begin{axis}[
    xlabel={Depth},
    ylabel={Expressibility},
    grid=major,
    legend style={at={(0.701,0.975)}, anchor=north, legend columns=2, font=\small},
      width=9.5cm,
      height=6cm,
    mark options={scale=0.8}
]

% Saturation annotations

%% 4 Qubits
\addplot+ [smooth, mark=diamond*] coordinates {
(1,0.2992588070702943) (3,0.2299310508131574) (4,0.1937978545786053) (5,0.1718677742922541)
(6,0.1323406531599291) (7,0.109214757106687) (8,0.0917542266354478) (9,0.0677738837163404)
(10,0.07309284991232) (11,0.0562471895954709) (12,0.0559962255915956) (13,0.0548576626438662)
(14,0.0537138899728284) (15,0.0491728892858171) (16,0.0520237963993616)
};

%% 6 Qubits
\addplot+ [smooth, mark=square*] coordinates {
(1,0.5072236601164235) (2,0.2295132038900902) (3,0.2241992113030118) (4,0.2162775540351582)
(5,0.2031109685701057) (6,0.1919327852260174) (7,0.1807002124017458) (8,0.1517991736168742)
(9,0.1316027648375149) (10,0.1302054297632877) (11,0.1080340030586106) (12,0.0960083495861189)
(13,0.0766062024405924) (14,0.0694802049853964) (15,0.0702767053155697) (16,0.0570785431945649)
(17,0.0588525101925812) (18,0.0552592247726873) (19,0.0512515521261127) (20,0.0504471027000516)
(21,0.0481046266411808) (22,0.0504417902161269) (23,0.0517825992374332) (24,0.0519914959864635)
};

%% 10 Qubits
\addplot+ [smooth, mark=pentagon*] coordinates {
(1,0.3534694595733021) (2,0.1447434842661689) (3,0.1129769699231324) (4,0.1095125312591348)
(5,0.1016256000871471) (6,0.1061179475396146) (7,0.1221540951688805) (8,0.1113881680472401)
(9,0.1059114740002213) (10,0.1055106818721966) (11,0.1375039714071025) (12,0.1189851404711129)
(13,0.1255957613960129) (14,0.1284836839091647) (15,0.1237884377625955) (16,0.0984407788231127)
(17,0.1013239762047458) (18,0.0989834786952364) (19,0.089539378837572) (20,0.0920875525102265)
(21,0.0912257217249331) (22,0.0699655208102525) (23,0.0695467713299103) (24,0.059883467035005)
(25,0.0572150429110123) (26,0.0634619311359408) (27,0.0554703773732048) (28,0.0538400864042016)
(29,0.053330384373083) (30,0.0530033998703314) (31,0.0531042836973888) (32,0.0500357352145293)
(33,0.0486741318193968) (34,0.0502558296677912) (35,0.0473171667568007) (36,0.0466798143112722)
(37,0.047372481566468) (38,0.0498408849869556) (39,0.0526630665737006) (40,0.048384179374624)
};

\addplot+ [smooth, mark=star ] coordinates {
(1,0.15063704864119873) (2,0.07286615147829153) (3,0.0482651005918967) (4,0.05742105601627835)
(5,0.04465238486912876) (6,0.048110806631813215) (7,0.04588969145854158) (8,0.04807905542638293)
(9,0.052099615558453605) (10,0.06511143705767663) (11,0.06422362942316468) (12,0.07077131221390781)
(13,0.07070426268454957) (14,0.07151494273934058) (15,0.07284925951062748) (16,0.07927651094949739)
(17,0.07989251065067334) (18,0.08074293219701823) (19,0.08528180384161501) (20,0.0869474486058778)
(21,0.0772986940664552) (22,0.08820408165102745) (23,0.08277853458430229) (24,0.07347553728616679)
(25,0.07002179348115272) (26,0.07460326424756521) (27,0.07201727477188744) (28,0.06282892162155433)
(29,0.05541282071653811) (30,0.06145953127617377) (31,0.05250664992329259) (32,0.05382335222432547)
(33,0.04987270796374435) (34,0.04863307500878922) (35,0.05041540590440834) (36,0.04622969729702487)
};

\legend{4 Qubits, 6 Qubits, 10 Qubits,  14 Qubits}

\end{axis}
\end{tikzpicture}
\caption{Expressibility saturation with increasing circuit depth. The plot shows the expressibility metric (y-axis) as a function of circuit depth (x-axis) for ansatze constructed using gate set A on systems of 4, 6, 10, and 14 qubits. The expressibility rapidly improves with initial increases in depth before reaching a plateau, which indicates that optimal expressibility is achieved at a circuit depth ( $\approx2n $ depth, where $n$ is the number of qubits) that scales linearly with the number of qubits.  }
\label{fig:exp_saturation}
\end{figure}

\subsubsection{Gate set}
\label{subsubsec:gate set}

The composition of the gates, single-qubit gates like (RX, RY, RZ, H) and two-qubit gates (CNOT) \cite{barenco1995elementary} is the primary factor influencing the GA's search for an optimal ansatz. We evaluated nine gate sets (A-I), detailed in Table \ref{table:gate_set_revised}, to understand this relationship. The resulting expressibility for each set is presented in Figure \ref{fig:exp_vs_gateset}, which reveals a performance gap across all test system sizes and gate sets. Gate sets A, B, and H consistently produce ansatze with higher expressibility. The distinguishing feature of these top-performing sets is a structural constraint of the CNOT gate to be applied only on adjacent qubits. \textcolor{black}{This locality constraint introduces more double and single-qubit gates, resulting in a high gate and parameter count.} The higher count of entangling operations of CNOT allows the ansatz to more effectively and uniformly explore Hilbert space, which is the direct cause of its enhanced expressibility score.

\textcolor{black}{In contrast, gate sets such as D, E, and I exhibit slightly lower expressibility compared to other gate sets while achieving this expressibility with significantly fewer gates and parameters. This improved resource efficiency leads to more favorable optimization landscapes, enhancing trainability and VQE performance.}

% We investigate these combinations of gates in the genetic algorithm framework and find that gate sets A, B, and H perform the best with the lowest expressibility values, but gate sets C, D, E, and G also perform well and are in the same range.

% \begin{tikzpicture}
% \begin{axis}[
%     ybar,
%     bar width=6pt,
%       width=9.5cm,
%       height=6cm,
%     enlarge x limits={abs=0.5cm},
%     ymin=0, ymax=70,
%     ylabel={Gate Count},
%     xlabel={Gate Set},
%     symbolic x coords={A,B,C,D,E,F,G,H,I},
%     xtick=data,
%     legend style={at={(0.9605,0.975
%     )},
%         anchor=north east,
%         cells={anchor=west},
%         font=\small,legend columns=3
%     },
%     grid=major,
%     grid style={dashed,gray!30},
%     bar shift=0pt,
%     cycle list={{fill=black},{fill=red},{fill=green}}
% ]

% % Parameterized gates
% \addplot+[bar shift=-6pt] coordinates {
%     (A,0.041539) (B,0.0418283) (C,0.049942) (D,0.056392)
%     (E,0.0458831) (F,0.050675) (G,0.047175) (H,0.0432459) (I,0.043029)
% };

% % Unparameterized gates
% \addplot+[bar shift=6pt] coordinates {
%     (A,0.042638) (B,0.05259) (C,0.06383) (D,0.06858)
%     (E,0.09509) (F,0.09107) (G,0.08357) (H,0.043396) (I,0.06719)
% };

% % Total gates (sum of parameterized and unparameterized)
% \addplot+[bar shift=0pt] coordinates {
%     (A,0.03959928) (B,0.0386732) (C,0.09133) (D,0.088)
%     (E,0.09066) (F,0.07904) (G,0.081151) (H,0.040409) (I,0.05212)
% };

% \legend{P Gates, U Gates, T Gates}

% \end{axis}
% \end{tikzpicture}

\begin{figure*}
    \centering
    \begin{tikzpicture}
    \begin{axis}[
        ybar,
        bar width=5pt,
        width=12.8cm,
        height=7.2cm,
        enlarge x limits={abs=1.2cm},
        ymin=0, ymax=0.12,
        ylabel={Expressibility},
        xlabel={Gate Set},
        ytick={0,0.02,0.04,0.06,0.08,0.10,0.12},
        symbolic x coords={A,B,C,D,E,F,G,H,I},
        xtick align=center,
        xtick=data,
        legend style={at={(0.015,0.972)},
            anchor=north west,
            cells={anchor=west},
            font=\large,
            draw=black,
            fill=none
        },
        grid=major,
        grid style={dashed,gray!30},
    ]

    % Colors: black, red, green, orange, violet
    \addplot+[ybar, bar shift=-10pt, fill=blue] coordinates {
        (A,0.041539) (B,0.0418283) (C,0.049942) (D,0.056392)
        (E,0.0458831) (F,0.050675) (G,0.047175) (H,0.0432459) (I,0.043029)
    };

    \addplot+[ybar, bar shift=-5pt, fill=red] coordinates {
        (A,0.0449418) (B,0.0426769) (C,0.053475) (D,0.053679)
        (E,0.058091) (F,0.058306) (G,0.071813) (H,0.0437248) (I,0.058367)
    };

    \addplot+[ybar, bar shift=0pt, fill=green!70!black] coordinates {
        (A,0.042638) (B,0.05259) (C,0.06383) (D,0.06858)
        (E,0.09509) (F,0.09107) (G,0.08357) (H,0.043396) (I,0.06719)
    };

    \addplot+[ybar, bar shift=5pt, fill=orange] coordinates {
        (A,0.037759) (B,0.043685) (C,0.088753) (D,0.07846)
        (E,0.1111922) (F,0.04931) (G,0.07856) (H,0.04249) (I,0.0492)
    };

    \addplot+[ybar, bar shift=10pt, fill=violet] coordinates {
        (A,0.03959928) (B,0.0386732) (C,0.09133) (D,0.088)
        (E,0.09066) (F,0.07904) (G,0.081151) (H,0.040409) (I,0.05212)
    };

    \legend{4 Qubits, 8 Qubits, 10 Qubits, 12 Qubits, 14 Qubits}

    \end{axis}
    \end{tikzpicture}
    \caption{The chart compares the best expressibility (y-axis) achieved by the GA framework for nine gate sets (x-axis) across various system sizes, with their compositions defined in Table \ref{table:gate_set_revised}. Gate sets A, B, and H consistently demonstrate superior performance. Their higher expressibility is linked to the adjacent-CNOT constraint, which results in the generation of circuits with a larger number of entangling gates.}
    \label{fig:exp_vs_gateset}
\end{figure*}

\subsubsection{Gate Count} \label{subsubsec:resource requirement}
To understand the resource cost associated with the expressibility of each gate set, we analyze the gate count composition of the resulting circuits. Fig. \ref{fig:gate_count} illustrates the breakdown of parametrized, unparametrized, and total gate counts for the ansatze evolved by our GA framework on a 4-qubit, depth-16 system. We find a direct correlation between high expressibility and gate count. Notably, gate sets with superior expressibility- A, B, and H also yield circuits with the highest total number of gates.  Conversely, gate set I is the most efficient, requiring the fewest total gates, but this efficiency comes at the cost of expressive power. This highlights a fundamental tradeoff in ansatz design: while a greater number of gates can improve expressiveness, it increases complexity and potential susceptibility to hindrance in trainability. Therefore, an optimal gate set must balance achieving sufficient expressibility with the minimum necessary gate count to ensure better trainability.

\begin{figure}
\centering
\begin{tikzpicture}
\begin{axis}[
    ybar,
    bar width=6pt,
      width=9.5cm,
      height=6cm,
    enlarge x limits={abs=0.5cm},
    ymin=0, ymax=70,
    ylabel={Gate Count},
    xlabel={Gate Set},
    symbolic x coords={A,B,C,D,E,F,G,H,I},
    xtick=data,
    legend style={at={(0.986,0.975
    )},
        anchor=north east,
        cells={anchor=west},
        font=\small,legend columns=3
    },
    grid=major,
    grid style={dashed, gray!30},
    bar shift=0pt,
    cycle list={{
    fill={rgb,255:red,55; green,126; blue,184}},   % Strong Blue
    {fill={rgb,255:red,228; green,26; blue,28}},   % Strong Red
    {fill={rgb,255:red,77; green,175; blue,74}}    % Strong Green
}
]

% Parameterized gates
\addplot+[bar shift=-6pt] coordinates {
    (A,29) (B,27) (C,27) (D,20) (E,27) (F,26) (G,22) (H,28) (I,24)
};

% Unparameterized gates
\addplot+[bar shift=6pt] coordinates {
    (A,20) (B,19) (C,16) (D,15) (E,7) (F,11) (G,13) (H,25) (I,8)
};

% Total gates (sum of parameterized and unparameterized)
\addplot+[bar shift=0pt] coordinates {
    (A,49) (B,46) (C,43) (D,35) (E,34) (F,37) (G,35) (H,53) (I,32)
};

\legend{P Gates, NP Gates, T Gates}

\end{axis}
\end{tikzpicture}
\caption{The bar plot shows the gate counts for GA desiginged ansatze on a 4-qubit, depth-16 circuit, categorized by gate set (x-axis). The bars represent the number of parametrized (P Gates), non-parametrized (NP Gates), and total (T Gates) gates. A strong correlation is observed between highly expressive gate sets (A, B, and H) and higher total gate counts, while more resource-efficient gate sets like E and I show lower expressibility.}
\label{fig:gate_count}
\end{figure}

\vspace{-0.5cm}
\subsubsection{Time taken for classical optimization} \label{subsubsec:classical runtime}
Beyond the quantum resources required, the classical computational cost of the ansatz search is a key consideration. Figure \ref{fig:cir_gen_time} shows the runtime of our GA framework, revealing that the optimization time scales with both the number of qubits and depth for all gate sets. This scaling with both qubit count and depth is further quantified in Table \ref{tab:execution-time-kn}. Gate sets also have an impact on the computational cost. However, the classical runtime should be viewed as a one-time investment rather than a recurring operational cost. Once the GA discovers an optimal circuit structure, that ansatz becomes a reusable tool for different Hamiltonian and VQA problems, without repeating the search process.  This approach effectively amortizes the classical overhead.

\begin{figure}
\centering
\begin{tikzpicture}
\begin{axis}[
    ybar,
    bar width=6pt,
     width=9.5cm,
      height=6cm,
    enlarge x limits={abs=0.8cm},
    ymin=0, ymax=800,
    ylabel={Time (s)},
    xlabel={Gate Set},
    symbolic x coords={A,B,C,D,E,F,H,I},
    xtick align=center,
    xtick=data,
    legend style={at={(0.775,0.97)},
        anchor=north east,
        cells={anchor=west},
        font=\fontsize{11pt}{13.2pt}\selectfont,legend columns=3
    },
    grid=major,
    grid style={dashed,gray!30},
    bar shift=0pt,
cycle list={{
    fill={rgb,255:red,0; green,114; blue,178}},   % Dark Blue
    {fill={rgb,255:red,213; green,94; blue,0}},   % Strong Orange
    {fill={rgb,255:red,204; green,121; blue,167}} % Magenta
}
]

% 2 Qubits
\addplot+[bar shift=-6pt] coordinates {
    (A,112.97) (B,96.11) (C,92.28) (D,93.60) (E,92.78) (F,95.95) (H,104.01) (I,150.15)
};

% 4 Qubits
\addplot+[bar shift=0pt] coordinates {
    (A,248.11) (B,205.21) (C,184.66) (D,173.96) (E,176.17) (F,187.45) (H,227.45) (I,434)
};

% 8 Qubits
\addplot+[bar shift=6pt] coordinates {
    (A,644.04) (B,594.58) (C,453.66) (D,411.63) (E,417.77) (F,477.28) (H,668.18) (I,734)
};

\legend{2 Qubits, 4 Qubits, 8 Qubits}

\end{axis}
\end{tikzpicture}
\caption{The classical execution time (y-axis) required to run the genetic algorithm for 10 generations is shown as a function of gate sets (x-axis) and system size (2, 4, and 8 qubits). The runtime scales with the number of qubits, the gate set, and depth.}
\label{fig:cir_gen_time}
\end{figure}

\subsubsection{Performance on VQE} \label{subsec:tfim}

After analyzing the structural properties of the ansatze, we evaluated their practical performance on the Variation Quantum Eigensolver (VQE). We used the Transverse-Field Ising Model (TFIM) as our test bed to assess how effectively each ansatz could find the ground-state energy of the Hamiltonian. Figure \ref{fig: tfi_hamilconvergence} plots the energy throughout the VQE optimization for an 8-qubit, 16-depth circuit. While most ansatze drive the energy toward the ground state, sets B, G, and H perform poorly, either converging slowly or getting trapped in local minima far from the true ground-state energy.
%From the VQE energy convergence results, we can see that on a 4-qubit 16-layer ansatz, Set C performs the worst and gets stuck in local minima, while Set A, B, E, F, and G converge to the actual ground state energy with minimal error.

\begin{figure*}[htbp]
\centering
\color{black}
\begin{tikzpicture}[font=\small]

\begin{axis}[
      width=12.8cm,
    height=7.2cm,        % Adjusted to reverse width/height
    xlabel={Iteration},
    ylabel={Energy (Hartree)},
    legend style={
        at={(0.99,0.98)},
        anchor=north east,
        cells={anchor=west},
        font=\footnotesize,
    },
    grid=both,
    grid style={line width=.1pt, draw=gray!10},
    major grid style={line width=.2pt,draw=gray!50},
    xmin=0,
    xmax=100,
    ymin=-10.2,
    ymax=4
    ,
    tick label style={font=\scriptsize},
    label style={font=\small},
    title style={font=\small}
]

% Exact energy line
\addplot[
    black,
    dashed,
    line width=1pt,
] coordinates {(0,-9.837951) (100,-9.837951)};
\addlegendentry{Exact Diagonalization} 

% Set A
\addplot[
    color=blue,
    solid,
    line width=1pt,
    mark=none
] table[x=step, y=energy, col sep=comma] {csv1/energy_set_A.csv};
\addlegendentry{Set A}

% Set B
\addplot[
    color=red,
    solid,
    line width=1pt,
    mark=none
] table[x=step, y=energy, col sep=comma] {csv1/energy_set_B.csv};
\addlegendentry{Set B}

% Set C
\addplot[
    color=green!60!black,
    solid,
    line width=1pt,
    mark=none
] table[x=step, y=energy, col sep=comma] {csv1/energy_set_C.csv};
\addlegendentry{Set C}

% Set D
\addplot[
    color=orange,
    solid,
    line width=1pt,
    mark=none
] table[x=step, y=energy, col sep=comma] {csv1/energy_set_D.csv};
\addlegendentry{Set D}

% Set E
\addplot[
    color=purple,
    solid,
    line width=1pt,
    mark=none
] table[x=step, y=energy, col sep=comma] {csv1/energy_set_E.csv};
\addlegendentry{Set E}

% Set F
\addplot[
    color=cyan!70!black,
    solid,
    line width=1pt,
    mark=none
] table[x=step, y=energy, col sep=comma] {csv1/energy_set_F.csv};
\addlegendentry{Set F}

% Set G
\addplot[
    color=brown,
    solid,
    line width=1pt,
    mark=none
] table[x=step, y=energy, col sep=comma] {csv1/energy_set_G.csv};
\addlegendentry{Set G}

% Set H
\addplot[
    color=magenta,
    solid,
    line width=1pt,
    mark=none
] table[x=step, y=energy, col sep=comma] {csv1/energy_set_H.csv};
\addlegendentry{Set H}

% Set I
\addplot[
    color=teal,
    solid,
    line width=1pt,
    mark=none
] table[x=step, y=energy, col sep=comma] {csv1/energy_set_I.csv};
\addlegendentry{Set I}

\end{axis}
\end{tikzpicture}

\caption{\textcolor{black}{The plot tracks the energy (y-axis) as a function of optimization iterations (x-axis) for an 8-qubit TFIM Hamiltonian with a 16-depth circuit.
The dashed line represents the analytically calculated ground-state energy.
While most gate sets converge accurately, ansatze from gate sets B, G, and H struggle to reach the ground state, highlighting trainability issues.}}
\label{fig: tfi_hamilconvergence}
\end{figure*}

\textcolor{black}{To assess the performance of GA ansatze, we analyze the energy error (percentage \%) after convergence across multiple system sizes, as shown in Figure \ref{fig:energy_errors_gatesets}. 
The plot shows that gate set E and A consistently achieve the lowest energy errors, demonstrating their superior performance in the VQE task.
These results lead to a crucial insight: high expressibility alone does not guarantee VQE success, as seen in the case of the gate sets B and H.
The superior performance of set E suggests that an optimal ansatz for VQE requires a balance between sufficient expressibility and a resource-efficient gate set that promotes better trainability. 
This shows us that ansatz design must properly balance expressibility and resource efficiency to perform well as a general-purpose ansatz.}

%Set E performs the best out of all the gate sets with minimal errors

\begin{figure}[htbp]
\centering
\color{black}
\begin{tikzpicture}
\begin{axis}[
    ybar=1pt,           % Adds 1pt space BETWEEN bars in the same group
    bar width=4pt,      % Reduced from 6pt to prevent group-to-group overlap
    width=9.5cm,         % Increased width to give more horizontal room
    height=7cm,
    enlarge x limits={abs=0.8cm}, % Provides padding at the far left and right
    ymin=0, ymax=60,
    ylabel={Energy Error \%},
    xlabel={Gate Set},
    symbolic x coords={A,B,C,D,E,F,G,H,I},
    xtick align=center,
    xtick=data,
    legend style={
        at={(0.5,0.85)}, % Lifted higher to avoid clashing with the top of the bars
        anchor=south,
        legend columns=4, 
        cells={anchor=west},
        font=\large
    },
    grid=major,
    grid style={dashed, gray!30},
    cycle list={
        {fill=blue!80, draw=black},    
        {fill=red!80, draw=black},      
        {fill=green!60!black, draw=black}, 
        {fill=orange!80, draw=black}    
    }
]

% 4-qubit
% \addplot+ coordinates {
%     (A,0.046726) (B,0.126000) (C,0.32158) (D,0.059965) (E,0.053562) (F,0.29367) (G,0.053579) (H,0.927130) (I,0.181452)
% };
\addplot+ coordinates {
    (A,0.981892)
    (B,2.647743)
    (C,6.757629)
    (D,1.260095)
    (E,1.125543)
    (F,6.171132)
    (G,1.125900)
    (H,19.482555)
    (I,3.813002)
};

% 8-qubit
\addplot+ coordinates {
    (A,4.803485)
    (B,16.775694)
    (C,7.089807)
    (D,0.314392)
    (E,0.245363)
    (F,0.229658)
    (G,47.956408)
    (H,27.682409)
    (I,2.216145)
};

% 10-qubit
\addplot+ coordinates {
    (A,3.408758)
    (B,11.022922)
    (C,0.336890)
    (D,5.388423)
    (E,1.688028)
    (F,7.970479)
    (G,21.391626)
    (H,49.651254)
    (I,1.150710)
};

% 12-qubit
\addplot+ coordinates {
    (A,0.162857)
    (B,11.942734)
    (C,0.628388)
    (D,3.550858)
    (E,0.254710)
    (F,5.502061)
    (G,18.697115)
    (H,0.934097)
    (I,8.047919)
};

\legend{4 qubit, 8 qubit, 10 qubit, 12 qubit}
\end{axis}
\end{tikzpicture}
\caption{\textcolor{black}{We test our GA ansatz on the TFIM with coupling ($J$) = 1 and transverse field ($h$) = 1 for different system size ranging from 4 qubits to 12 qubits. Energy errors for each gate set is reported. We see a consistent trend where most gate sets converge near to the ground state energy except B, G and H.}}
\label{fig:energy_errors_gatesets}
\end{figure}

\subsection{Molecular Hamiltonian benchmarking:}
\label{subsec:mol}

\textcolor{black}{To test the robustness and practical performance of our GA-designed ansatz, we benchmarked it to approximate molecular ground-state energies using VQE. We consider four standard molecular systems, H$_2$, LiH, BeH$_2$, and H$_2$O, simulated within the STO-3g minimal basis set. Figure \ref{Molecular Hamiltonian} illustrates the convergence behavior of the GA ansatz, demonstrating stable and system coverage towards the exact diagonalization reference energy across all molecular systems considered.}

To place this performance in context, Table \ref{Table II} compares the final energies obtained using the GA ansatz with those achieved by two widely used approaches:  the chemistry-inspired UUCSD and the adaptive ADAPT-VQE. The results clearly show that the GA ansatz performs competitively, yielding final energies comparable to these established methods. While chemical accuracy is achieved for the smallest system (H$_2$), for larger molecules the GA–VQE energies remain very close to the Exact Diagonalization values, with deviations on the order of $10^{-2}$ Hartree, despite the restricted circuit depth.

Notably, this accuracy is obtained with a reduced circuit depth. For LiH, BeH$_2$, and H$_2$O, the GA ansatz requires depths of at most 48, whereas UCCSD depths exceed 1400 and ADAPT-VQE depths scale in the order of $200-300$. This substantial reduction in circuit depth indicates that the GA ansatz is resource-efficient and scalable. These results demonstrate that the GA framework can generate reusable, problem-agnostic ansatze that achieve systematic convergence and competitive accuracy while significantly reducing circuit complexity.

\begin{figure*}[ht]

\centering
\color{black}
\begin{tikzpicture}
% Define shared properties for all plots in the group
\begin{groupplot}[
    group style={
        group size=2 by 2,
        horizontal sep=2cm,
        vertical sep=2.5cm
    },
    xlabel={Iterations},
    ylabel={Energy (Hartree)},
    grid=major,
    width=0.4\textwidth,
    height=0.3\textwidth,
    /tikz/mark options={scale=0.5} 
]

% --- Plot 1 (Top-Left) ---
% Style the legend specifically for this plot
\nextgroupplot[
    title={H$_2$},
    legend style={
        font=\medium,
        at={(0.95,0.95)},
        anchor=north east
    }
]
\addplot [dashed, gray, thick] coordinates { (0, -1.13619) (150, -1.13619)};
\addplot+ [smooth, mark=*, color=blue] coordinates {
    (0, -0.1) (1, -0.06) (2, -0.5) (3, -0.43) (4, -0.57) (5, -0.65) (6, -0.68) (7, -0.69) (8, -0.67) (9, -0.67) (10, -0.7) (11, -0.74) (12, -0.76) (13, -0.77) (14, -0.77) (15, -0.78) (16, -0.77) (17, -0.78) (18, -0.8) (19, -0.81) (20, -0.82) (21, -0.83) (22, -0.85) (23, -0.88) (24, -0.91) (25, -0.95) (26, -1.01) (27, -1.06) (28, -1.09) (29, -1.08) (30, -1.07) (31, -1.07) (32, -1.07) (33, -1.08) (34, -1.08) (35, -1.08) (36, -1.09) (37, -1.1) (38, -1.11) (39, -1.11) (40, -1.12) (41, -1.12) (42, -1.12) (43, -1.12) (44, -1.12) (45, -1.12) (46, -1.12) (47, -1.12) (48, -1.13) (49, -1.13) (50, -1.12) (51, -1.13) (52, -1.13) (53, -1.13) (54, -1.13) (55, -1.13) (56, -1.13) (57, -1.13) (58, -1.13) (59, -1.13) (60, -1.13) (61, -1.13) (62, -1.13) (63, -1.13) (64, -1.13) (65, -1.13) (66, -1.13) (67, -1.13) (68, -1.13) (69, -1.13) (70, -1.14) (71, -1.14) (72, -1.14) (73, -1.14) (74, -1.14) (75, -1.14) (76, -1.14) (77, -1.14) (78, -1.14) (79, -1.14) (80, -1.14) (81, -1.14) (82, -1.14) (83, -1.14) (84, -1.14) (85, -1.14) (86, -1.14) (87, -1.14) (88, -1.14) (89, -1.14) (90, -1.14) (91, -1.14) (92, -1.14) (93, -1.14) (94, -1.14) (95, -1.14) (96, -1.14) (97, -1.14) (98, -1.14) (99, -1.14) (100, -1.14) (101, -1.14) (102, -1.14) (103, -1.14) (104, -1.14) (105, -1.14) (106, -1.14) (107, -1.14) (108, -1.14) (109, -1.14) (110, -1.14) (111, -1.14) (112, -1.14) (113, -1.14) (114, -1.14) (115, -1.14) (116, -1.14) (117, -1.14) (118, -1.14) (119, -1.14) (120, -1.14) (121, -1.14) (122, -1.14) (123, -1.14) (124, -1.14) (125, -1.14) (126, -1.14) (127, -1.14) (128, -1.14) (129, -1.14) (130, -1.14) (131, -1.14) (132, -1.14) (133, -1.13) (134, -1.13) (135, -1.13) (136, -1.12) (137, -1.12) (138, -1.13) (139, -1.13) (140, -1.13) (141, -1.13) (142, -1.14) (143, -1.13) (144, -1.13) (145, -1.14) (146, -1.13) (147, -1.13) (148, -1.13) (149, -1.14) (150, -1.13)
};
\legend{Exact Diagonalization, GA Ansatz}

% --- Plot 2 (Top-Right) ---
% Style the legend specifically for this plot
\nextgroupplot[
    legend style={
        title={BeH$_2$},
        font=\medium,
        at={(0.95,0.95)},
        anchor=north east
    }
]
\addplot [dashed, gray, thick] coordinates { (0, -14.878) (150, -14.878) };
\addplot+ [smooth, mark=square*, color=red] coordinates {
    (0, -6.58) (1, -7.38) (2, -6.04) (3, -7.72) (4, -7.51) (5, -8.14) (6, -9.13) (7, -10.49) (8, -10.82) (9, -11.64) (10, -12.27) (11, -12.35) (12, -12.85) (13, -12.94) (14, -13.2) (15, -13.43) (16, -13.47) (17, -13.7) (18, -13.76) (19, -13.79) (20, -13.89) (21, -13.93) (22, -14.07) (23, -14.17) (24, -14.21) (25, -14.26) (26, -14.28) (27, -14.35) (28, -14.36) (29, -14.4) (30, -14.43) (31, -14.48) (32, -14.5) (33, -14.57) (34, -14.55) (35, -14.55) (36, -14.57) (37, -14.6) (38, -14.62) (39, -14.62) (40, -14.63) (41, -14.64) (42, -14.64) (43, -14.68) (44, -14.71) (45, -14.7) (46, -14.39) (47, -14.71) (48, -14.53) (49, -14.72) (50, -14.58) (51, -14.7) (52, -14.68) (53, -14.64) (54, -14.7) (55, -14.66) (56, -14.72) (57, -14.7) (58, -14.71) (59, -14.76) (60, -14.75) (61, -14.78) (62, -14.77) (63, -14.77) (64, -14.79) (65, -14.79) (66, -14.8) (67, -14.82) (68, -14.82) (69, -14.81) (70, -14.82) (71, -14.84) (72, -14.83) (73, -14.83) (74, -14.84) (75, -14.84) (76, -14.84) (77, -14.84) (78, -14.84) (79, -14.84) (80, -14.85) (81, -14.85) (82, -14.84) (83, -14.83) (84, -14.82) (85, -14.81) (86, -14.77) (87, -14.76) (88, -14.76) (89, -14.73) (90, -14.7) (91, -14.77) (92, -14.83) (93, -14.79) (94, -14.8) (95, -14.82) (96, -14.77) (97, -14.75) (98, -14.7) (99, -14.57) (100, -14.66) (101, -14.77) (102, -14.78) (103, -14.75) (104, -14.77) (105, -14.79) (106, -14.78) (107, -14.83) (108, -14.78) (109, -14.8) (110, -14.81) (111, -14.76) (112, -14.8) (113, -14.81) (114, -14.84) (115, -14.81) (116, -14.81) (117, -14.81) (118, -14.81) (119, -14.79) (120, -14.79) (121, -14.83) (122, -14.83) (123, -14.83) (124, -14.84) (125, -14.83) (126, -14.85) (127, -14.85) (128, -14.84) (129, -14.83) (130, -14.81) (131, -14.81) (132, -14.8) (133, -14.8) (134, -14.81) (135, -14.83) (136, -14.81) (137, -14.82) (138, -14.84) (139, -14.82) (140, -14.81) (141, -14.82) (142, -14.85) (143, -14.83) (144, -14.84) (145, -14.83) (146, -14.83) (147, -14.81) (148, -14.82) (149, -14.83) (150, -14.83)
};
\legend{Exact Diagonalization, GA Ansatz}

% --- Plot 3 (Bottom-Left) ---
% Style the legend specifically for this plot
\nextgroupplot[
    legend style={
        title={LiH},
        font=\medium,
        at={(0.95,0.95)},
        anchor=north east
    }
]
\addplot [dashed, gray, thick] coordinates { (0, -7.6799) (150, -7.6799) };
\addplot+ [smooth, mark=triangle*, color=green!60!black] coordinates {
    (0, -2.57) (1, -3.42) (2, -4.94) (3, -6.02) (4, -5.93) (5, -6.21) (6, -6.57) (7, -6.73) (8, -6.87) (9, -7.04) (10, -7.07) (11, -7.13) (12, -7.1) (13, -7.22) (14, -7.2) (15, -7.24) (16, -7.28) (17, -7.34) (18, -7.43) (19, -7.36) (20, -7.35) (21, -7.34) (22, -7.37) (23, -7.44) (24, -7.43) (25, -7.46) (26, -7.44) (27, -7.44) (28, -7.45) (29, -7.46) (30, -7.49) (31, -7.49) (32, -7.5) (33, -7.49) (34, -7.48) (35, -7.49) (36, -7.49) (37, -7.51) (38, -7.51) (39, -7.51) (40, -7.51) (41, -7.51) (42, -7.51) (43, -7.51) (44, -7.52) (45, -7.52) (46, -7.52) (47, -7.51) (48, -7.52) (49, -7.52) (50, -7.52) (51, -7.52) (52, -7.52) (53, -7.52) (54, -7.52) (55, -7.52) (56, -7.52) (57, -7.53) (58, -7.52) (59, -7.53) (60, -7.52) (61, -7.53) (62, -7.53) (63, -7.53) (64, -7.53) (65, -7.53) (66, -7.53) (67, -7.53) (68, -7.54) (69, -7.54) (70, -7.54) (71, -7.55) (72, -7.55) (73, -7.56) (74, -7.56) (75, -7.57) (76, -7.57) (77, -7.58) (78, -7.58) (79, -7.59) (80, -7.59) (81, -7.59) (82, -7.59) (83, -7.59) (84, -7.59) (85, -7.59) (86, -7.59) (87, -7.59) (88, -7.59) (89, -7.59) (90, -7.6) (91, -7.63) (92, -7.65) (93, -7.64) (94, -7.6) (95, -7.62) (96, -7.64) (97, -7.64) (98, -7.65) (99, -7.63) (100, -7.64) (101, -7.64) (102, -7.65) (103, -7.64) (104, -7.66) (105, -7.65) (106, -7.65) (107, -7.65) (108, -7.66) (109, -7.65) (110, -7.65) (111, -7.65) (112, -7.66) (113, -7.66) (114, -7.66) (115, -7.66) (116, -7.66) (117, -7.66) (118, -7.66) (119, -7.66) (120, -7.66) (121, -7.66) (122, -7.66) (123, -7.66) (124, -7.66) (125, -7.66) (126, -7.66) (127, -7.66) (128, -7.66) (129, -7.66) (130, -7.66) (131, -7.66) (132, -7.65) (133, -7.64) (134, -7.64) (135, -7.65) (136, -7.66) (137, -7.65) (138, -7.65) (139, -7.66) (140, -7.65) (141, -7.66) (142, -7.66) (143, -7.66) (144, -7.66) (145, -7.66) (146, -7.66) (147, -7.66) (148, -7.66) (149, -7.66)
};
\legend{Exact Diagonalization, GA Ansatz}

% --- Plot 4 (Bottom-Right) ---
% Style the legend specifically for this plot
\nextgroupplot[
    legend style={
        title={H$_2$O},
        font=\medium,
        at={(0.95,0.95)},
        anchor=north east
    }
]
\addplot [dashed, gray, thick] coordinates { (0, -73.226) (149, -73.226) };
\addplot+ [smooth, mark=diamond*, color=orange] coordinates {
    (0, -64.04) (1, -66.7) (2, -67.64) (3, -68.43) (4, -68.54) (5, -69.36) (6, -69.76) (7, -70.1) (8, -70.48) (9, -70.79) (10, -70.81) (11, -70.96) (12, -71.52) (13, -71.83) (14, -71.84) (15, -71.92) (16, -72.05) (17, -72.33) (18, -72.52) (19, -72.66) (20, -72.77) (21, -72.68) (22, -72.68) (23, -72.63) (24, -72.68) (25, -72.89) (26, -72.91) (27, -72.92) (28, -72.96) (29, -72.94) (30, -73.02) (31, -73.05) (32, -73.04) (33, -73.03) (34, -73.03) (35, -73.1) (36, -73.13) (37, -73.12) (38, -73.12) (39, -73.11) (40, -73.13) (41, -73.14) (42, -73.16) (43, -73.17) (44, -73.16) (45, -73.18) (46, -73.17) (47, -73.17) (48, -73.17) (49, -73.18) (50, -73.19) (51, -73.2) (52, -73.19) (53, -73.19) (54, -73.19) (55, -73.19) (56, -73.2) (57, -73.2) (58, -73.2) (59, -73.2) (60, -73.2) (61, -73.2) (62, -73.2) (63, -73.21) (64, -73.21) (65, -73.21) (66, -73.21) (67, -73.21) (68, -73.21) (69, -73.21) (70, -73.21) (71, -73.21) (72, -73.21) (73, -73.21) (74, -73.21) (75, -73.21) (76, -73.21) (77, -73.21) (78, -73.21) (79, -73.21) (80, -73.21) (81, -73.21) (82, -73.21) (83, -73.21) (84, -73.21) (85, -73.21) (86, -73.21) (87, -73.21) (88, -73.21) (89, -73.21) (90, -73.21) (91, -73.21) (92, -73.21) (93, -73.21) (94, -73.21) (95, -73.21) (96, -73.21) (97, -73.21) (98, -73.21) (99, -73.21) (100, -73.21) (101, -73.21) (102, -73.21) (103, -73.21) (104, -73.21) (105, -73.21) (106, -73.21) (107, -73.21) (108, -73.21) (109, -73.21) (110, -73.21) (111, -73.21) (112, -73.21) (113, -73.21) (114, -73.21) (115, -73.21) (116, -73.21) (117, -73.21) (118, -73.21) (119, -73.21) (120, -73.21) (121, -73.21) (122, -73.21) (123, -73.21) (124, -73.21) (125, -73.21) (126, -73.21) (127, -73.21) (128, -73.21) (129, -73.21) (130, -73.21) (131, -73.21) (132, -73.21) (133, -73.21) (134, -73.21) (135, -73.21) (136, -73.21) (137, -73.21) (138, -73.21) (139, -73.21) (140, -73.21) (141, -73.21) (142, -73.21) (143, -73.21) (144, -73.21) (145, -73.21) (146, -73.21) (147, -73.21) (148, -73.21) (149, -73.2)
};
\legend{Exact Diagonalization, GA Ansatz}
\end{groupplot}
\end{tikzpicture}
\caption{VQE convergence of the GA ansatz on Molecular Hamiltonians. VQE optimization for GA-designed ansatz on four benchmark molecules, H$_2$, LiH, BeH$_2$, and H$_2$O. Each plot shows the calculated energy (y-axis) converging to the exact ground state energy (dashed line) as a function of optimization iterations (x-axis). The rapid convergence across all four systems demonstrates the effectiveness and general applicability of the GA ansatz for quantum chemistry problems.}
\label{Molecular Hamiltonian}
\end{figure*}

\textcolor{black}{The final energy values of all three ansatze are compared in Table~\ref{Table II}}.

\begin{table}[t]
\color{black}
\centering
\caption{\textcolor{black}{Performance benchmark against UCCSD and ADAPT-VQE. 
 Comparison of final ground-state energies (Hartree) and circuit depths for different molecular systems using the GA-designed ansatz, UCCSD, ADAPT-VQE, and Exact diagonalization.}}
\footnotesize
\renewcommand{\arraystretch}{1.1}
\setlength{\tabcolsep}{4pt}

\begin{tabular}{lccccc}
\hline\hline
\textbf{Mol} & \textbf{Metric} & \textbf{GA} & \textbf{UCCSD} & \textbf{ADAPT} & \textbf{Exact Diag.} \\
\hline
\multirow{2}{*}{H$_2$}
& Energy & $-1.13608$ & $-1.13608$ & $-1.13608$ & $-1.13608$ \\
\cline{2-6}
& Depth  & 12 & 20 & 24 & -- \\

\hline
\multirow{2}{*}{LiH}
& Energy & $-7.86266$ & $-7.8817$ & $-7.8822$ & $-7.88266$ \\
\cline{2-6}
& Depth  & 48 & 1423 & 208 & -- \\

\hline
\multirow{2}{*}{BeH$_2$}
& Energy & $-14.84842$ & $-14.854$ & $-14.876$ & $-14.87803$ \\
\cline{2-6}
& Depth  & 48 & 1423 & 192 & -- \\

\hline
\multirow{2}{*}{H$_2$O}
& Energy & $-73.212$ & $-73.22$ & $-73.2244$ & $-73.226$ \\
\cline{2-6}
& Depth  & 48 & 1442 & 284 & -- \\
\hline\hline
\end{tabular}

\label{Table II}
\end{table}

\vspace{-0.5cm}

\subsection{\textcolor{black}{Noisy simulations}}
\label{subsec:noisy_simulations}

\textcolor{black}{To assess the performance of the ansatz produced by the genetic algorithm and compare it to ADAPT under noisy conditions, we repeat the H$_2$ simulation for both ansatze in the presence of noise.
We use a noise model derived from a Qiskit backend with $N_{\mathrm{shots}}=1000$. 
The noise model is constructed from error data generated and sampled randomly from historical IBM backend data and incorporated into PennyLane as shown in their tutorial}~\cite{UtkarshAzad2024}.

\textcolor{black}{We change the optimizer to SPSA (gradient-free) for the GA ansatz, as gradient-free optimizers perform better in noisy conditions. 
For H$_2$ we observe that the GA circuits have a lower depth than the corresponding ADAPT circuits which translates into better noise resilience and lower mean final energy closer to the true ground-state energy, as we can see from the Table~\ref{tab:noisy_ga_adapt_tab}.}

\begin{figure}[t]
    \centering
    \color{black}
    \begin{tikzpicture}
\begin{axis}[
      width=0.4\textwidth,
      height=0.3\textwidth,
    xlabel={Iteration},
    ylabel={Energy (Hartree)},
    legend style={
        at={(0.95,0.95)},
        anchor=north east,
        cells={anchor=west},
        font=\medium,
    },
    grid=both,
    grid style={line width=.1pt, draw=gray!10},
    major grid style={line width=.2pt, draw=gray!50},
    xmin=-50,
    xmax=655,
    ymin=-1.35,
    ymax=1,
    tick label style={font=\scriptsize},
    label style={font=\small},
    title style={font=\small},
]

% Exact energy line
\addplot[
    gray,
    dashed,
    line width=1pt,
] coordinates {(0,-1.13608) (605,-1.13608)};
\addlegendentry{Exact Diagonalization}

% GA Ansatz (CSV, index-based)
\addplot[
    color=teal,
    solid,
    line width=1pt,
    mark=none
] table[
    x expr=\coordindex,
    y index=0
] {csv1/noisy.csv};
\addlegendentry{GA Ansatz}

\end{axis}
\end{tikzpicture}

    \caption{\textcolor{black}{Convergence plot for the Genetic Ansatz optimization on H$_2$ under noisy conditions.}}
    \label{fig:ga_noisy_convergence}
\end{figure}

% \begin{figure}[t]
%     \centering
%     \color{black}
%     \includegraphics[width=0.85\linewidth]{Images/noisy_conv.png}
%     \caption{\textcolor{black}{Convergence plot for the Genetic Ansatz optimization on H$_2$ under noisy conditions.}}
%     \label{fig:ga_noisy_convergence}
% \end{figure}

\begin{table}[t]
    \centering
    \color{black}
    \caption{\textcolor{black}{Comparison of noisy simulation results for ADAPT-VQE and the GA evolved ansatz on the H$_2$ Hamiltonian. SEM stands for standard error of mean, and the lowest observed absolute energy error is with respect to the exact ground-state energy obtained from diagonalization of the H$_2$ Hamiltonian.}}
    \label{tab:noisy_ga_adapt_tab}
    \begin{tabular}{lccc}
        \hline
        Ansatz & Mean energy & SEM & Lowest energy error \\
        \hline
        ADAPT 
        & $-1.1414401$ 
        & $4.90 \times 10^{-3}$ 
        & $4.40 \times 10^{-3}$ \\
        GA ansatz 
        & $-1.1352861$ 
        & $1.92 \times 10^{-4}$ 
        & $4.17 \times 10^{-4}$ \\
        \hline
    \end{tabular}
\end{table}

\vspace{-0.5cm}
% \subsection{Data Availability:}
% The quantum circuits produced by our Genetic Algorithm (GA) are provided as supplementary material \cite{Mallapur_Genetic_Ansatz_2025}. Access to this repository is available upon request. 
% Interested researchers may contact the authors for access.

 \subsection{Data Availability}

 The quantum circuit architectures evolved by the Genetic Algorithm (GA), along with the implementation of the genetic framework, are publicly available in the repository associated with Ref.~\cite{Mallapur_Genetic_Ansatz_2025}.
Additional information or extended data related to the study can be obtained from the authors upon reasonable request.

\vspace{-0.5cm}
\subsection{Discussion}\label{sec7}
In this work, we demonstrate that a genetic-algorithm-based framework can generate expressive, resource-efficient ansatze that are reusable across different Hamiltonians, eliminating the need to design a new circuit for each system. 
By evolving a single ansatz and then directly applying it with only classical parameter optimization (for both molecular and non-molecular systems), we achieve several key strengths.

First, the reusability of our approach is demonstrated by the fact that the same evolved circuit structure was applied successfully to LiH, BeH$_2$, H$_2$O, and the transverse-field Ising model\textcolor{black}{(12 qubit (Set E))}, converging within 1\% of the exact ground-state energies without re-running the genetic search. Second, our method exhibits competitive performance: across all benchmarks, our GA-derived ansatze perform \textcolor{black}{almost on par with the problem-specific designs such as UCCSD and ADAPT-VQE.} Notably, Gate Set E consistently yielded the lowest final energy errors, and our circuits match UCCSD's accuracy while using $O(n^2)$ parametrized gates rather than $O(n^3)$ for UCCSD. \textcolor{black}{Third, the evolved circuits demonstrate resource efficiency by maintaining shallow depths and minimal gate counts.} Finally, our approach provides reduced classical overhead since once the ansatz is generated, only classical optimizations are needed for new problems, saving the significant classical runtime otherwise spent on repeated architecture searches.

\textcolor{black}{Therefore, we have obtained problem-agnostic circuits that balance expressibility and resource efficiency, perform robustly on various Hamiltonians, and greatly simplify the VQA workflow on NISQ devices.}

\vspace{-0.5cm}
\section{Conclusion}\label{sec8}
\textcolor{black}{Designing effective ansatze for variational quantum algorithms requires balancing expressibility and resource efficiency}. Circuits must be able to represent target quantum states while remaining optimizable on noisy intermediate-scale quantum devices. Our genetic algorithm framework addresses this challenge by systematically evolving quantum circuit architectures that maximize expressibility within fixed depth constraints.

\textcolor{black}{Our results demonstrate that genetically optimized ansatze achieve performance comparable to established methods like UCCSD and ADAPT-VQE across multiple molecular systems, while using fewer parameterized gates.} The key advancement lies in the reusability of evolved circuits: once generated, these ansatze can be applied across different Hamiltonians with only classical parameter optimization, reducing the computational overhead of repeated circuit design for new quantum problems.

\textcolor{black}{With this work, we provide an ansatz design algorithm for variational quantum algorithms, where a single evolved circuit architecture serves multiple quantum simulation tasks without sacrificing accuracy while being resource efficient. The demonstrated transferability of our approach suggests that genetic optimization represents a viable strategy for developing practical ansatz for variational quantum algorithms on near-term quantum hardware.}
% \vspace{-0.5cm}
\begin{acknowledgments}
We are grateful to the High Performance Computing (HPC) facility at IISER Bhopal, where large-scale calculations in this project were run. We thank Mainak Bhattacharyya for his valuable insights and the members of the QuCIS Lab for their constructive discussions throughout the course of this project.
The authors thank the funding received from the Department of Science and Technology under the National Quantum Mission. 
\end{acknowledgments}

\bibliography{bibliography}
\onecolumngrid
\appendix
\section*{Appendix}

\begin{figure*}[htbp]
    \centering
    \includegraphics[width=0.65\linewidth]{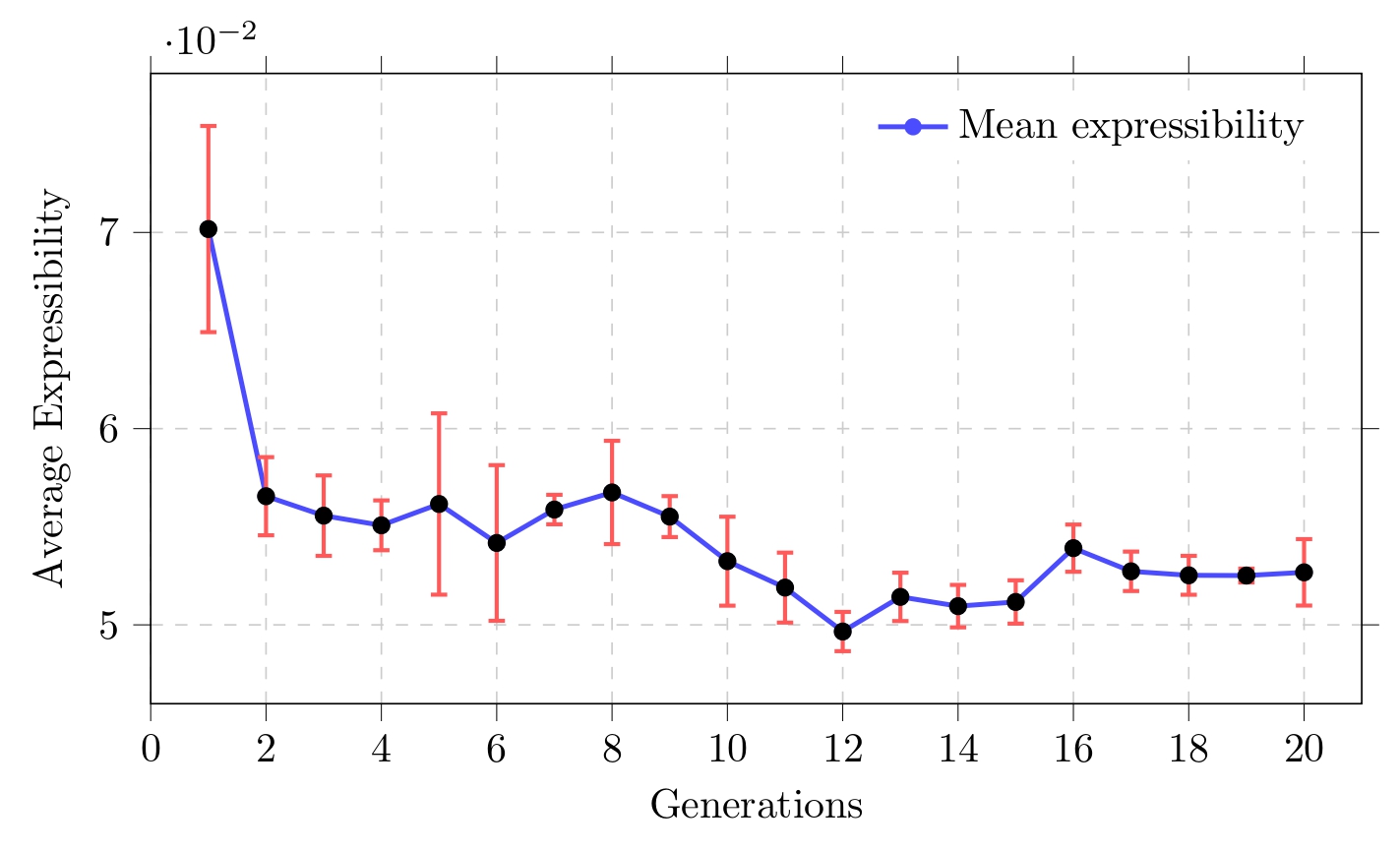}
    \caption{Genetic ansatz evolution over 20 generations over four different seeds. Extending the GA to 20 generations yields no significant improvement in expressibility beyond generation 10, indicating that additional generations only increase computational cost.}
    \label{fig:20-gen}
\end{figure*}

This appendix provides supplementary information to support the main text. Appendix \ref{app:GAFull} details the pseudocode for our genetic algorithm framework along with information on how long it takes to generate different circuits in table \ref{tab:execution-time-kn} and Figure \ref{fig:20-gen} showing the genetic algorithm framework being run for 20 generations. Appendix \ref{app:MolHam} provides the details for replicating the molecular Hamiltonians used for our experiment. In Appendix \ref{app:evolvedAnsatze}, we have provided some of the ansatze that we have used in our experiments.

% This appendix provides supplementary information to support the main text. Appendix \ref{secA1} details the pseudocode for our fidelity-based method for calculating the expressibility of a parametrized quantum circuit. Appendix \ref{secA2} provides further pseudocode outlining the core components of our GA-framework-based ansatz search, including population initialization, parent selection, and offspring generation.

% Finally, Appendix \ref{appendix:secA3} contains the full set of pairwise energy landscape plots that were generated for our trainability analysis.

\section{Genetic Algorithm Framework}
\label{app:GAFull}
This section provides pseudo-code for all the steps involved in our genetic algorithm framework for ansatz search. Algorithm \ref{alg:express} details the pseudocode for our fidelity-based method for calculating the expressibility of a parametrized quantum circuit. Further pseudocodes outline the core components of our GA-framework-based ansatz search, including population initialization(algorithm \ref{alg:init}), parent selection and offspring generation(algorithm \ref{alg:cross}), and population update after each generation(Algorithm \ref{alg:update}).

\begin{algorithm}[H]
\DontPrintSemicolon
\caption{Calculating Expressibility of a PQC}\label{alg:express}
\KwIn{$n$ = \# qubits, $S$ = \# samples}
\KwOut{$\mathcal{E}$ = Expressibility}
Define parametrized circuit on $n$ qubits\;
Initialize \texttt{fidelities} list\;
\For{$i\leftarrow1$ \KwTo $S$}{
  Sample $\theta_1,\theta_2\sim\mathcal{U}[0,2\pi]$\;
  $\psi_1\leftarrow$ circuit$(\theta_1)$,\quad $\psi_2\leftarrow$ circuit$(\theta_2)$\;
  $F\leftarrow|\langle\psi_1|\psi_2\rangle|^2$\;
  Append $F$ to \texttt{fidelities}\;
}
Build histogram of \texttt{fidelities} (bin size $b$)\;
Compute Haar distribution over the same bins\;
Normalize both to probability vectors\;
$\mathcal{E}\leftarrow D_{\mathrm{KL}}(\texttt{fidelities}\,\Vert\,\texttt{Haar})$\;
\Return{$\mathcal{E}$}
\end{algorithm}

% \subsection{Genetic algorithm based ansatz search}\label{secA2}

\begin{algorithm}[H]
\DontPrintSemicolon
\caption{Initialize Population and Evaluate Expressibility}\label{alg:init}
\KwIn{$n$ (qubits), $L$ (layers), $P$ (pop.\ size), $S$ (samples)}
\KwOut{Expressibility scores for each of $P$ circuits}
$\mathcal{G}\leftarrow\{\mathrm{RX,RY,RZ,CX,CRX,H}\}$\;
\For{$i\leftarrow1$ \KwTo $P$}{
  Randomly build circuit$_i$ with $L$ gates from $\mathcal{G}$\;
  Compute $\mathcal{E}_i\leftarrow$ Expressibility(circuit$_i$, $S$) (Alg.~\ref{alg:express})\;
}
\end{algorithm}

\begin{algorithm}[H]
\DontPrintSemicolon
\caption{Select Parents and Generate Offspring}\label{alg:cross}
\KwIn{Population of circuits + their $\mathcal{E}$}
\KwOut{Offspring population}
Sort circuits by ascending $\mathcal{E}$\;
Select top $k$ as parents\;
\For {\texttt{each offspring slot}}{
  Pick two parents $p_1,p_2$\;
  Choose crossover layer $\ell\in[1..L]$\;
  offspring $\leftarrow$ layers[1:\,$\ell$] of $p_1$ + layers[$\ell{+}1$:\,$L$] of $p_2$\;
  Assign new random parameters\;
  \If{\texttt{Uniform(0,1)}$<m$}{Mutate one gate or parameter\;}
  Compute offspring’s $\mathcal{E}$ (Alg.~\ref{alg:express})\;
}
\end{algorithm}

\vspace{-0.5pt}

\begin{algorithm}[H]
\DontPrintSemicolon
\caption{Update Population and Return Best Ansatz}\label{alg:update}
\KwIn{Current population, Offspring population}
\KwOut{Best ansatz found}

Replace population with offspring\;

Record best expressibility $\mathcal{E}$ of this generation\;

\If{\texttt{not last generation}}{
    Go to Alg~\ref{alg:cross}
}
% \Return{Ansatz with lowest $\mathcal{E}$}
\Return{\textnormal{Ansatz with lowest $\mathcal{E}$}}
\end{algorithm}

\begin{table}[h]
\centering
\caption{Average time (in seconds) to generate variational quantum circuits for different numbers of qubits $n$ and circuit depths $d = kn$, where $k \in \{1, 2, 3, 4\}$. Each entry shows the average time taken by the ansatz generation algorithm for given $n$ and $d$.} 
\label{tab:execution-time-kn}
\begin{tabular}{|c|c|c|c|c|}
\hline
\textbf{Qubits} & \textbf{$d=n$} & \textbf{$d=2n$} & \textbf{$d=3n$} & \textbf{$d=4n$} \\
\textbf{$n$} & \textbf{(s)} & \textbf{(s)} & \textbf{(s)} & \textbf{(s)} \\
\hline
2 & 24.257 & 32.877 & 46.744 & 52.203 \\
\hline
3 & 40.426 & 70.095 & 87.764 & 107.556 \\
\hline
4 & 65.067 & 101.094 & 143.742 & 186.507 \\
\hline
5 & 87.083 & 149.503 & 208.842 & 269.538 \\
\hline
6 & 115.716 & 214.969 & 304.193 & 400.754 \\
\hline
7 & 146.054 & 289.292 & 503.385 & 659.768 \\
\hline
8 & 241.702 & 726.306 & 1184.670 & 1563.758 \\
\hline
9 & 559.752 & 619.953 & 884.411 & 1190.944 \\
\hline
10 & 722.713 & 1460.488 & 2113.906 & 2767.386 \\
\hline
\end{tabular}
\end{table}

% \subsection{Ansatz evolution}

\section{Molecular Hamiltonians}
\label{app:MolHam}

Table \ref{tab:molecular_geometries} summarizes the molecular geometries and active-space definitions used to construct the chemical Hamiltonians. All Hamiltonians are generated using the STO-3G basis set. 
For BeH$_2$ and H$_2$O, a frozen-core approximation is employed, where the lowest-energy core orbital (corresponding to two core electrons) is frozen, and only the remaining valence orbitals are included in the active space. 
All other molecules are treated without freezing any core orbital.

\begin{table*}[htbp]
\centering
\caption{Molecular geometries and frozen-core settings used for constructing the chemical Hamiltonians. All distances are in \AA{} and angles in degrees.}
\label{tab:molecular_geometries}
\renewcommand{\arraystretch}{1.1}
\setlength{\tabcolsep}{6pt}
\begin{tabular}{|c|c|c|c|c|}
\hline
\textbf{Molecule} &
\textbf{Geometry} &
\textbf{Bond length(s) / Angle} &
\textbf{Frozen core} &
\textbf{Atomic coordinates (\AA)} \\
\hline
H$_2$ &
Linear &
H--H = $$1.32$$ &
None &
\begin{tabular}[c]{@{}l@{}}
H : $$(0, 0, -0.66)$$ \\
H : $$(0, 0, +0.66)$$
\end{tabular} \\
\hline
LiH &
Linear &
Li--H = $$2.9693$$ &
None &
\begin{tabular}[c]{@{}l@{}}
Li : $$(0, 0, 0)$$ \\
H  : $$(0, 0, 2.9693)$$
\end{tabular} \\
\hline
BeH$_2$ &
Linear &
Be--H = $$1.3264$$ &
$$2$$ e$^{-}$, $$1$$ orb. &
\begin{tabular}[c]{@{}l@{}}
Be : $$(0, 0, 0)$$ \\
H  : $$(0, 0, +1.3264)$$ \\
H  : $$(0, 0, -1.3264)$$
\end{tabular} \\
\hline
H$_2$O &
Bent &
\begin{tabular}[c]{@{}l@{}}
O--H = $$0.957$$ \\
$\angle$HOH = $$104.52$$
\end{tabular} &
$$2$$ e$^{-}$, $$1$$ orb. &
\begin{tabular}[c]{@{}l@{}}
O : $$(0, 0, 0)$$ \\
H : $$(+0.757, 0, 0.586)$$ \\
H : $$(-0.757, 0, 0.586)$$
\end{tabular} \\
\hline
\end{tabular}
\end{table*}

\vspace{3cm}

\section{Evolved Genetic Ansatze}
\label{app:evolvedAnsatze}
In this section, we present a subset of quantum circuits constructed with 8 qubits and a circuit depth of 16 for each considered gate set. These circuits are used for benchmarking in various instances throughout the paper.

% \begin{figure}[h]
%     \centering

    % \begin{subfigure}{0.48\columnwidth}
    %     \centering
    %     \includegraphics[width=\linewidth]{responses/responses/circuits/a.png}
    %     \caption*{Gate Set A}
    % \end{subfigure}\hfill
    % \begin{subfigure}{0.48\columnwidth}
    %     \centering
    %     \includegraphics[width=\linewidth]{responses/responses/circuits/b.png}
    %     \caption*{Gate Set B}
    % \end{subfigure}

    % \vspace{0.3cm}

    % \begin{subfigure}{0.48\columnwidth}
    %     \centering
    %     \includegraphics[width=\linewidth]{responses/responses/circuits/c.png}
    %     \caption*{Gate Set C}
    % \end{subfigure}\hfill
    % \begin{subfigure}{0.48\columnwidth}
    %     \centering
    %     \includegraphics[width=\linewidth]{responses/responses/circuits/d.png}
    %     \caption*{Gate Set D}
    % \end{subfigure}
    \begin{figure}[htbp]
        \centering
        \begin{minipage}[t]{0.48\columnwidth}
        \includegraphics[width=\linewidth]{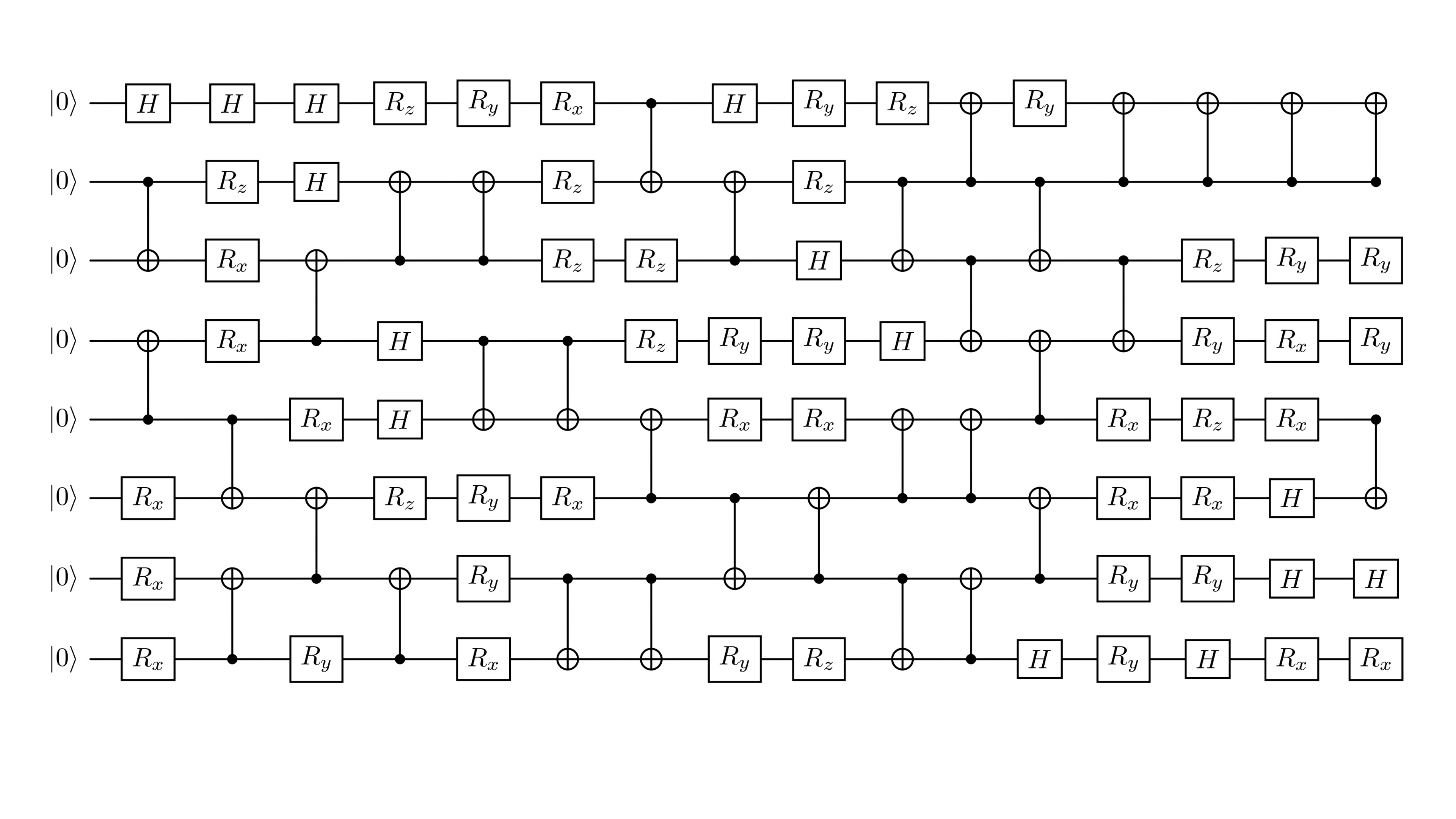}
        % \vspace{-0.2cm}
        \caption{Gate Set A}
    \end{minipage}
    \hfill
    \begin{minipage}[t]{0.48\columnwidth}
        \includegraphics[width=\linewidth]{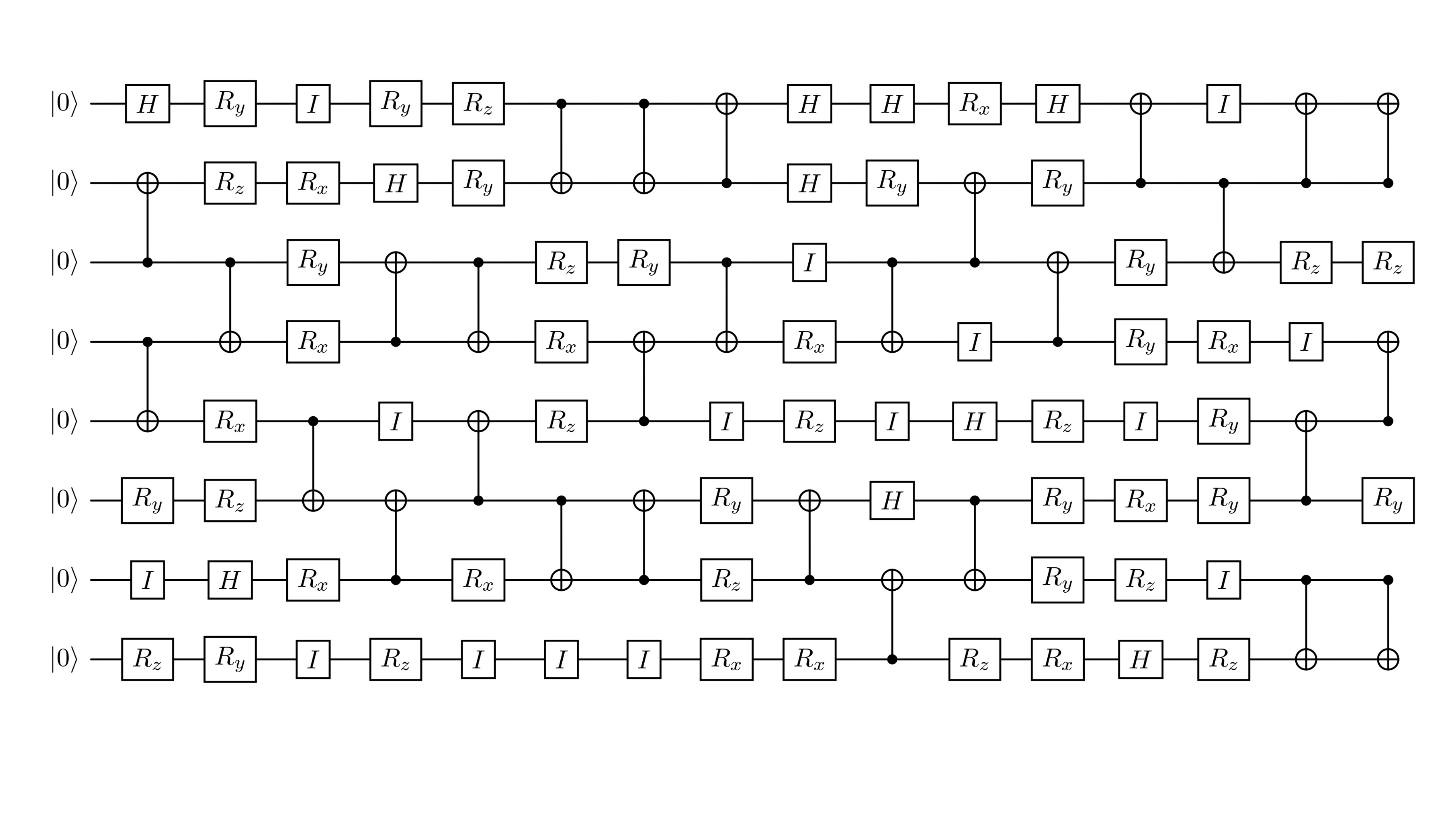}
        % \vspace{-0.2cm}
        \caption{Gate Set B}
        \vspace{0.5cm}
    \end{minipage}
     
    % \end{figure}
    % \begin{figure}[t]

    % \begin{minipage}[t]{0.48\columnwidth}
    %     \includegraphics[width=\linewidth]{Images/circuits/a.png}
    %     % \vspace{-0.2cm}
    %     \caption{Gate Set A}
    % \end{minipage}
    % \hfill
    % \begin{minipage}[t]{0.48\columnwidth}
    %     \includegraphics[width=\linewidth]{Images/circuits/b.png}
    %     % \vspace{-0.2cm}
    %     \caption{Gate Set B}
    %     \vspace{0.5cm}
    % \end{minipage}
     
   \begin{minipage}[t]{0.48\columnwidth}
        \includegraphics[width=\linewidth]{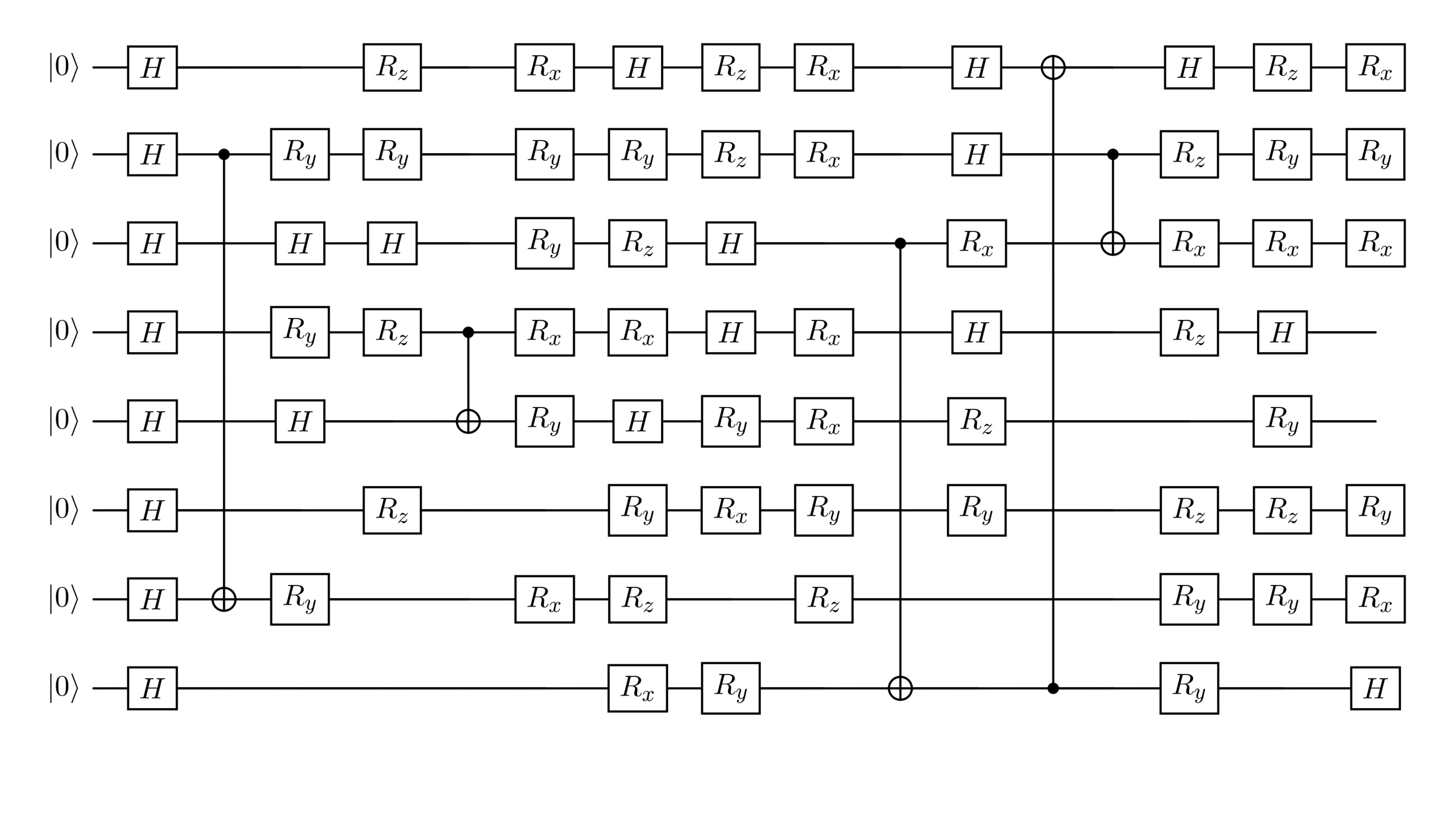}
        \caption{Gate Set C}
    \end{minipage} \hfill
    \begin{minipage}[t]{0.48\columnwidth}
        \includegraphics[width=\linewidth]{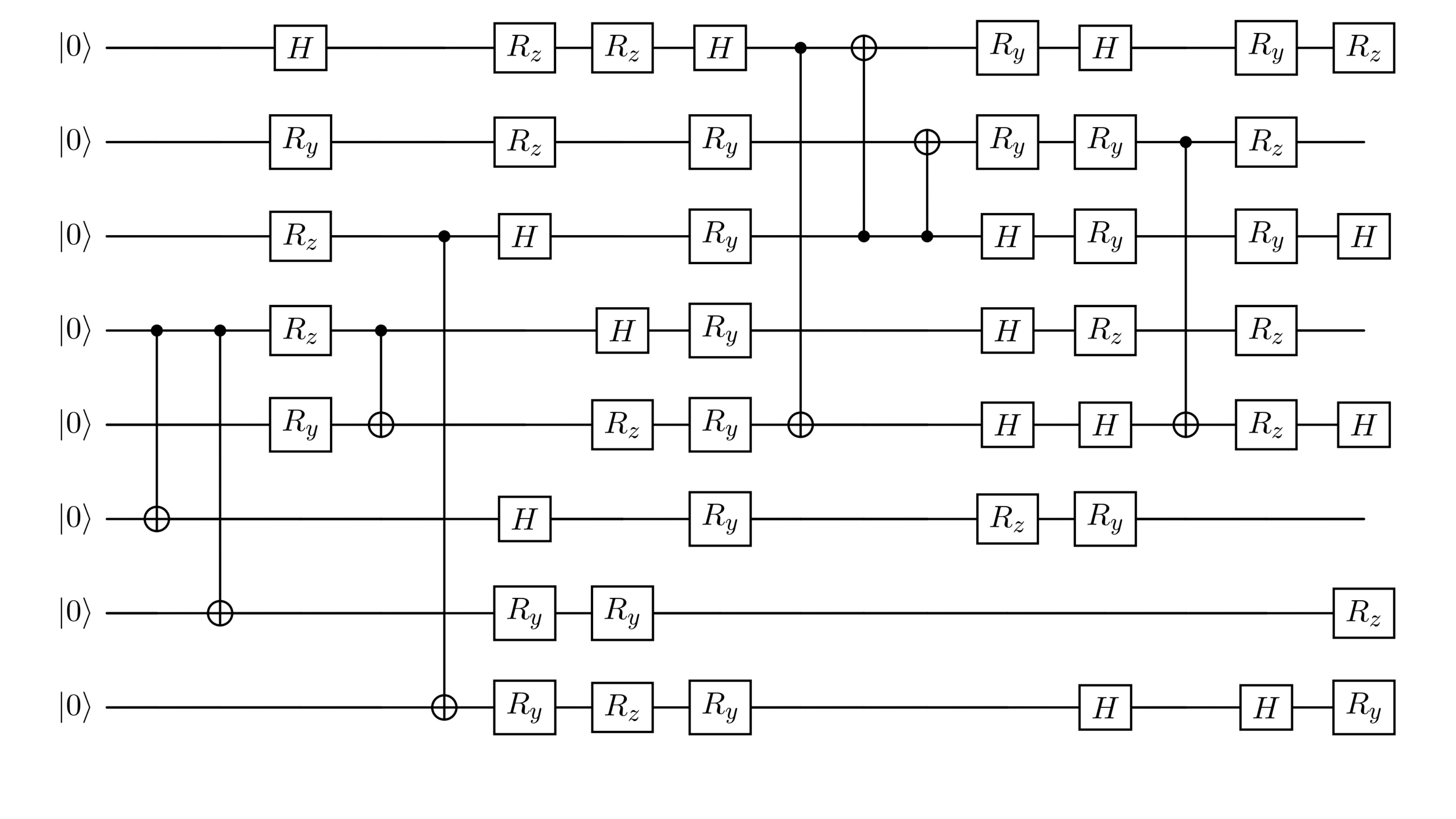}
        
        \caption{Gate Set D}
        \vspace{0.5cm}
    \end{minipage}
    \label{fig:ga_ansatz}

    \begin{minipage}[t]{0.48\columnwidth}
        \includegraphics[width=\linewidth]{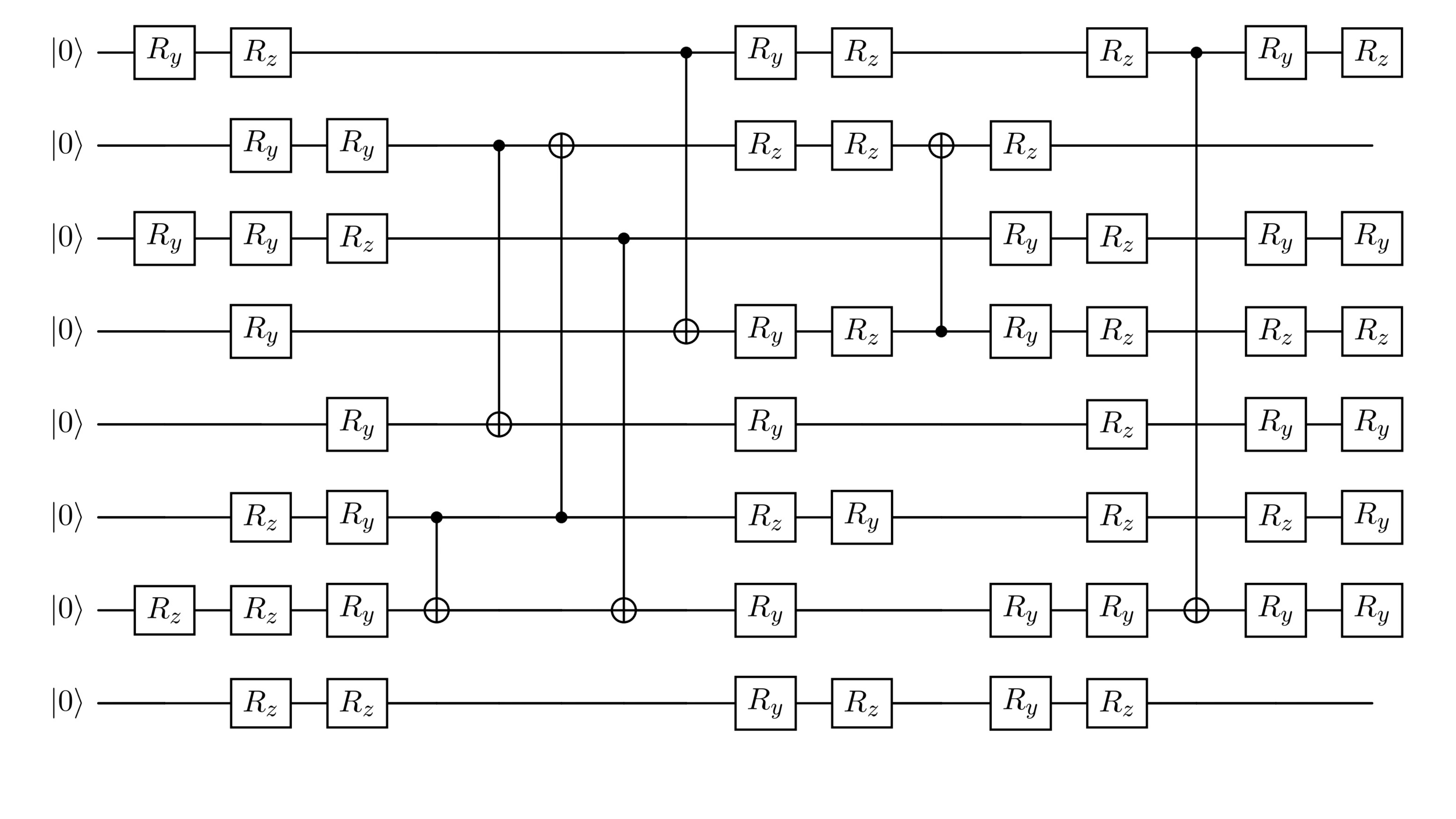}
        \caption{Gate Set E}
    \end{minipage}\hfill
    \begin{minipage}[t]{0.48\columnwidth}
        \includegraphics[width=\linewidth]{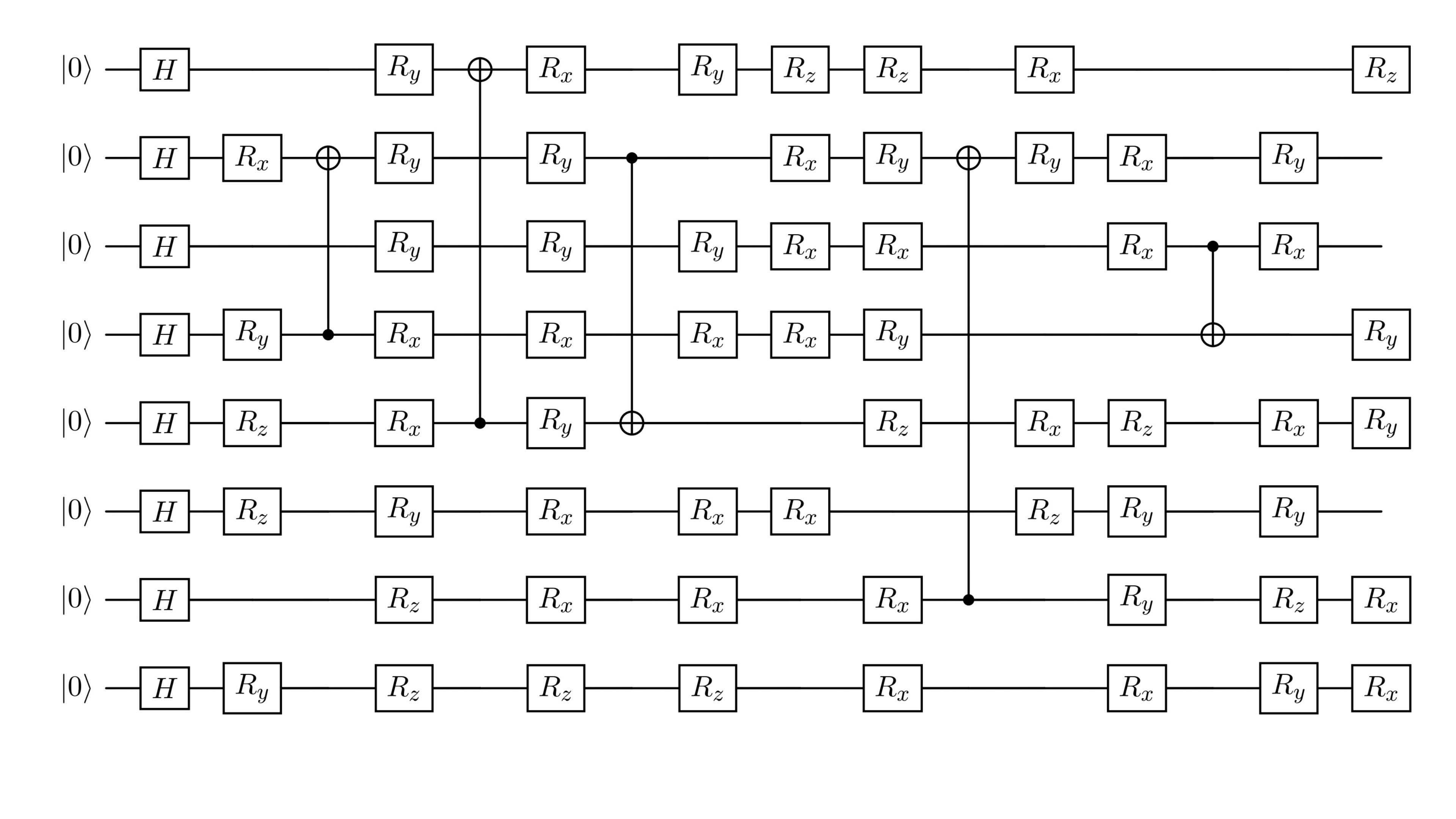}
        \caption{Gate Set F}
        \vspace{0.5cm}
    \end{minipage}
    
   \begin{minipage}[t]{0.48\columnwidth}
        \includegraphics[width=\linewidth]{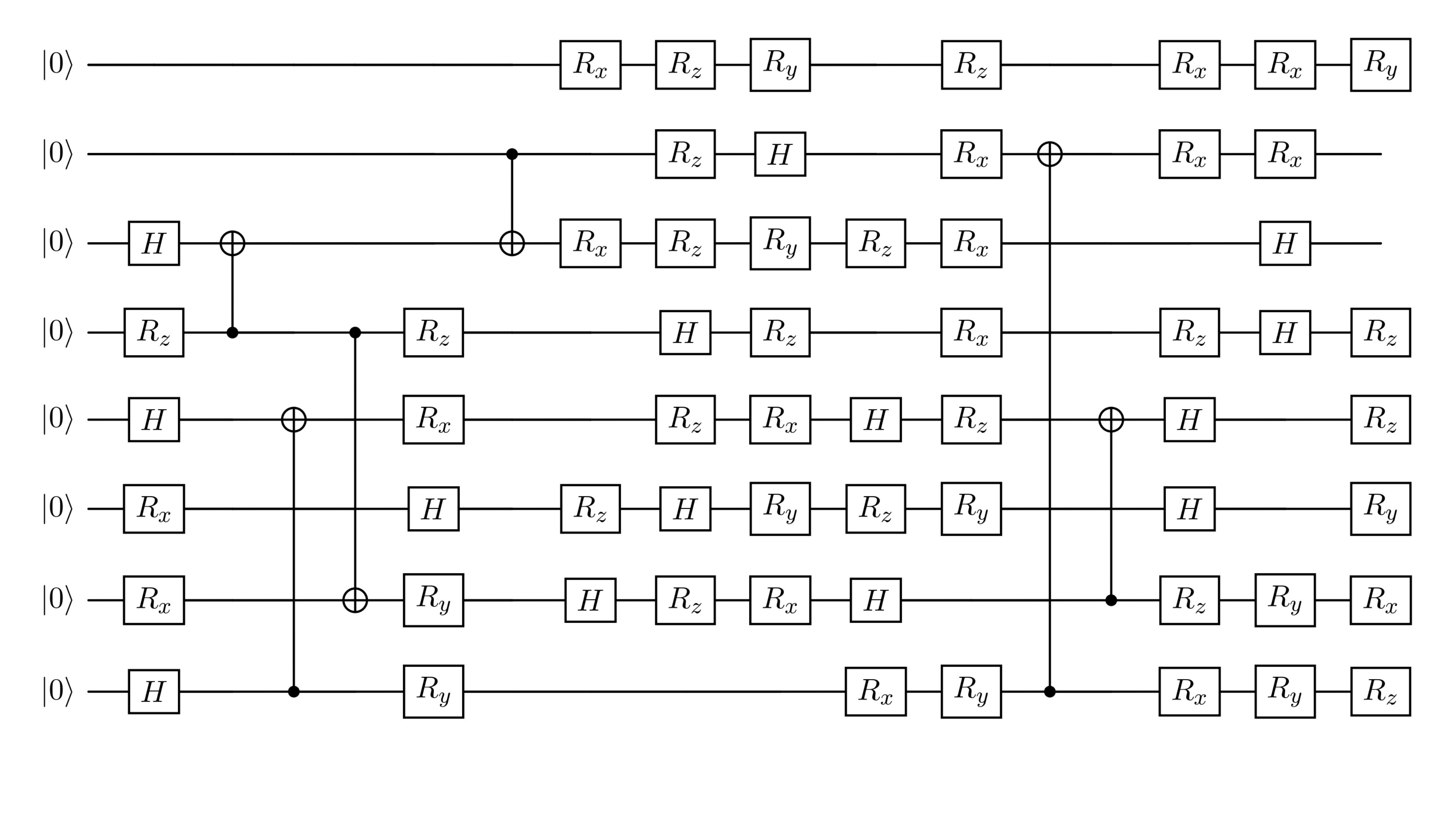}
        \caption{Gate Set G}
    \end{minipage}\hfill
    \begin{minipage}[t]{0.48\columnwidth}
        \includegraphics[width=\linewidth]{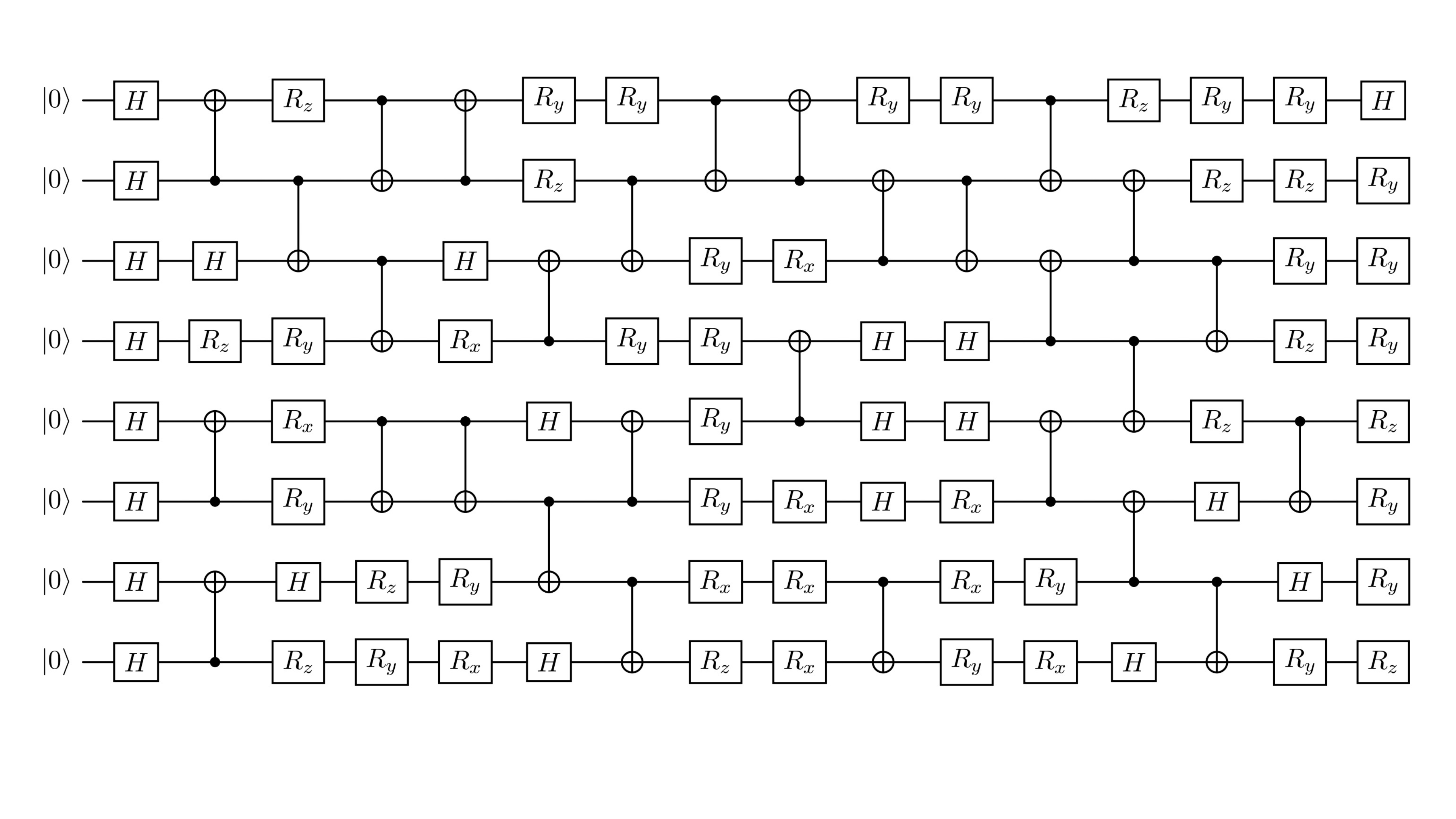}
        \caption{Gate Set H}
    \end{minipage}
    \label{fig:ga_ansatz}
\end{figure}

%     \label{fig:ga_ansatz_part1}
% \end{figure}

% \begin{figure}[t]
%     \centering

%     \begin{subfigure}{0.48\columnwidth}
%         \centering
%         \includegraphics[width=\linewidth]{responses/responses/circuits/e.png}
%         \caption*{Gate Set E}
%     \end{subfigure}\hfill
%     \begin{subfigure}{0.48\columnwidth}
%         \centering
%         \includegraphics[width=\linewidth]{responses/responses/circuits/f.png}
%         \caption*{Gate Set F}
%     \end{subfigure}

%     \vspace{0.3cm}

%     \begin{subfigure}{0.48\columnwidth}
%         \centering
%         \includegraphics[width=\linewidth]{responses/responses/circuits/g.png}
%         \caption*{Gate Set G}
%     \end{subfigure}\hfill
%     \begin{subfigure}{0.48\columnwidth}
%         \centering
%         \includegraphics[width=\linewidth]{responses/responses/circuits/h.png}
%         \caption*{Gate Set H}
%     \end{subfigure}

%     \label{fig:ga_ansatz_part2}
% \end{figure}

\begin{figure}[t]
    \centering
    \includegraphics[width=0.48\linewidth]{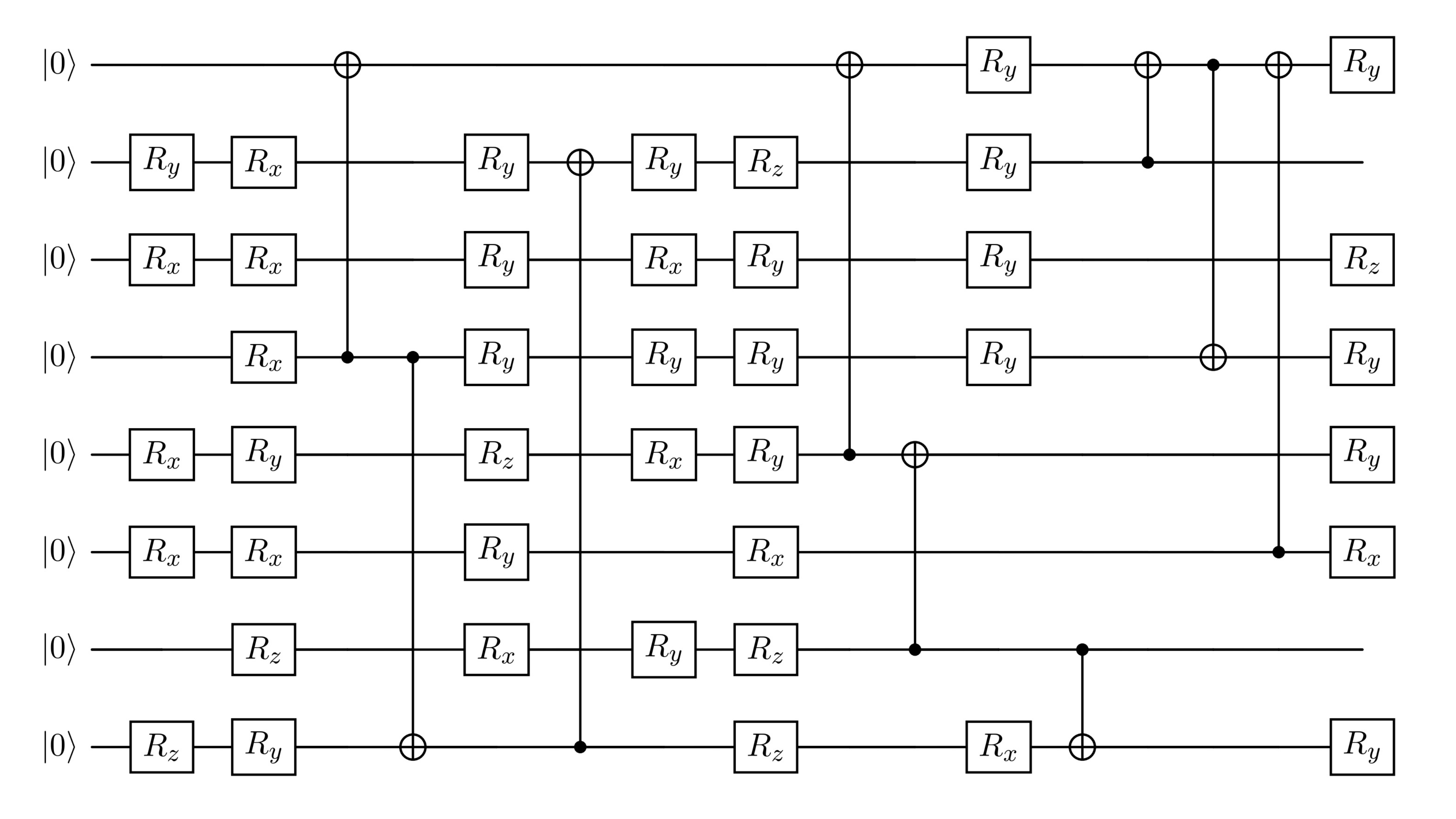}
    \caption{Gate Set I}
    \label{fig:ga_ansatz_part3}
\end{figure}

% Produces the bibliography via BibTeX.

\end{document}